\documentclass[fleqn,usenatbib]{mnras}

\usepackage[utf8]{inputenc}

\usepackage{acronym}
\usepackage{ae,aecompl}
\usepackage{booktabs}
\usepackage{dcolumn}
\usepackage{graphicx}
\usepackage{amsmath,amsfonts,amssymb}
\usepackage{mathrsfs}
\usepackage[normalem]{ulem}
\usepackage{epstopdf}

\usepackage[T1]{fontenc}

\let\oldhref\href
\renewcommand{\href}[2]{\oldhref{#1}{\hbox{#2}}}

\title[Neutrino shock acceleration in CCSN]{Non-thermal neutrinos created by shock acceleration in successful and failed core-collapse supernova}

\author[H. Nagakura and K. Hotokezaka]{
Hiroki Nagakura$^{1}$\thanks{E-mail: hirokin@astro.princeton.edu},
Kenta Hotokezaka$^{2,3}$,
\\
$^{1}$Department of Astrophysical Sciences, Princeton University, 4 Ivy Lane, Princeton, NJ 08544, USA \\
$^{2}$Research Center for the Early Universe, Graduate School of Science, University of Tokyo, Bunkyo-ku, Tokyo 113-0033, Japan\\
$^{3}$Kavli IPMU (WPI), UTIAS, The University of Tokyo, Kashiwa, Chiba 277-8583, Japan
}

\date{Accepted XXX. Received YYY; in original form ZZZ}

\pubyear{2020}

\begin{document}
\label{firstpage}
\pagerange{\pageref{firstpage}--\pageref{lastpage}}
\maketitle

\begin{abstract}
We present a comprehensive study of neutrino shock acceleration in core-collapse supernova (CCSN). The leading players are heavy leptonic neutrinos, $\nu_{\mu}$ and $\nu_{\tau}$; the former and latter potentially gain the energy up to $\sim 100$ MeV and $\sim 200$ MeV, respectively, through the shock acceleration. Demonstrating the neutrino shock acceleration by Monte Carlo neutrino transport, we make a statement that it commonly occurs in the early post bounce phase ($\lesssim 50$ ms after bounce) for all massive stellar collapse experiencing nuclear bounce and would reoccur in the late phase ($\gtrsim 100$ ms) for failed CCSNe. This opens up a new possibility to detect high energy neutrinos by terrestrial detectors from Galactic CCSNe; hence, we estimate the event counts for Hyper(Super)-Kamiokande, DUNE, and JUNO. We find that the event count with the energy of $\gtrsim 80$ MeV is a few orders of magnitude higher than that of the thermal neutrinos regardless of the detectors, and muon production may also happen in these detectors by $\nu_{\mu}$ with the energy of $\gtrsim 100$ MeV. The neutrino signals provide a precious information on deciphering the inner dynamics of CCSN and placing a constraint on the physics of neutrino oscillation; indeed, the detection of the high energy neutrinos through charged current reaction channels will be a smoking gun evidence of neutrino flavor conversion.
\end{abstract}


\section{Introduction}\label{sec:intro} 
Great effort has been dedicated over the past decades to study the mechanism of core-collapse supernova (CCSN) \citep[see recent reviews, e.g., ][]{2015PASA...32....9F,2016NCimR..39....1M,2017hsn..book.1095J,2017hsn..book.1575J,2018JPhG...45d3002H,2018SSRv..214...33B,2019ARNPS..6901918M,2020arXiv201009013M}. The canonical theory of driving the explosion is the neutrino-heating mechanism aided by multi-dimensional fluid instabilities. The thermal energy of proto-neutron star (PNS) is radiated and spatially transported by neutrinos, and then a fraction of the energy is converted into the post-shock matter by neutrino absorptions and non-isoenergetic scatterings, meanwhile the turbulent pressure is also increased due to the vigorous development of multi-dimensional fluid instabilities behind the shock; the non-linear coupling results in pushing the shock wave forward \citep[see, e.g.,][]{2019ApJ...887...43M,2019MNRAS.490.4622N,2020MNRAS.491.2715B,2020ApJ...890..127C}. If they are sufficient to prevail over accretion ram pressure, the shock wave transits into a runaway expansion phase, which marks successful explosions. In some peculiar cases, the PNS is supposed to collapse to a black hole (BH) if the mass accretion rate continues at a high rate even after shock revival \citep[fall-back mechanism; see, e.g.,][]{1994ApJ...423..659B,2018ApJ...855L...3O,Pan:2017tpk}, or if the shock revival is failed due to extremely high mass accretion rate \citep[failed CCSN, see, e.g.,][]{1999ApJ...524..262M,2001ApJ...550..410M,2006PhRvL..97i1101S,2011ApJ...730...70O,2011ApJ...737....6S,2020PhRvD.101l3013W}. Those scenarios are, however, conceived by purely theoretical arguments, indicating that they are still provisional. Comprehensive understanding of the CCSN dynamics requires the observational evidence that will be brought by neutrinos \citep{2010PhRvL.104y1101H,2013ApJS..205....2N,2018MNRAS.480.4710S,2019ApJ...881..139S,2021MNRAS.500..696N} and gravitational waves \citep{2016ApJ...829L..14K,2018ApJ...861...10M,2019MNRAS.486.2238A,2019MNRAS.487.1178P,2019ApJ...876L...9R,2019PhRvD.100h3008S,2020PhRvD.102b3028S,2020PhRvD.102b3027M}, and their joint analysis \citep{2016MNRAS.461.3296N,2018MNRAS.475L..91T,2019MNRAS.489.2227V,2020ApJ...898..139W,2020arXiv201003882S}.

The detection of neutrino signals is promising if CCSNe occur in our Galaxy. The history assures the feasibility; a breakthrough in the study of CCSN was made by the detections of neutrinos from SN 1987A occurred in the Large Magellanic Could \citep{IMB1987,kamioka,BNO1988}, in which the hypothesis that CCSNe are associated with the formation of a neutron star was proven. Over the last three decades, the scale and sensitivity of neutrino detectors have been expanded \citep[see, e.g.,][]{2012ARNPS..62...81S,2018JPhG...45d3002H}; for instance, Super-Kamiokande (SK) is a water-Cherenkov detector in operation, which has a capability of detecting ten thousands of neutrinos from Galactic CCSN \citep{2007ApJ...669..519I}. In several years, larger-scale detectors will be available; Hyper-Kamiokande (HK) is a scaled-up detector from SK, which will be available from 2027 and its fiducial volume will be several times larger than SK \citep{2018arXiv180504163H}. There are other future-planed neutrino detectors with different detection techniques; the Jiangmen Underground Neutrino Observatory (JUNO) \citep{2016JPhG...43c0401A}, a 20 kton liquid scintillator detector, will be soon in operation from 2021; the Deep Underground Neutrino Experiment, DUNE \citep{2016arXiv160105471A,2016arXiv160807853A,2020arXiv200806647A}, is another future-planned one, in which $40$ kton liquid-argon time-projection-chamber will be implemented. The coincidental neutrino detection by multiple detectors will shed light on the detailed features of CCSN neutrinos signals \citep{2021MNRAS.500..319N}. As such, the international network of these neutrino experiments is crucial for the analysis of CCSN neutrinos and their theoretical connection will be also tightened accordingly \citep[see, e.g., SNEWS collaboration,][]{2004NJPh....6..114A}.

It is interesting to consider how we take advantage of a large number of neutrino event counts ($> 1000$) to extract the information on the inner dynamics of CCSN. The time structure in neutrino signals is one of the targets, which will be resolved with a time scale of a few tens of milliseconds \citep[see, e.g.,][]{2018MNRAS.480.4710S}. Since the time evolution of the mass accretion rate in CCSN is imprinted in the neutrino signal, the timing of shock revival, progenitor structure, and BH formation \citep[if happens, see, e.g.,][]{2011ApJ...737....6S,2020PhRvD.101l3013W} would be constrained. Another interesting possibility is that the detailed feature of the energy spectrum of neutrinos can be investigated with high statistics data. The spectrum can be extracted by applying some statistical methods \citep{2002PhLB..547...37B,2002PhLB..542..239M,2008JCAP...12..006M,2017JCAP...11..036G,2014PhRvD..89f3007L,2016PhRvD..94b3006L,2018JCAP...04..040G,2018PhRvD..97b3019N} or direct spectrum unfolding techniques \citep{2019PhRvD..99l3009L,2021MNRAS.500..696N,2021MNRAS.500..319N} to the observed data. The flavor-dependent feature in the spectrum will be useful to prove matter state of the CCSN core and also clue to study neutrino flavor conversion.

Most previous studies have paid attention to the energy region with a few tens of MeV \citep[but see][]{2014PhRvD..89d3012M}, which makes sense, since neutrinos with the energy of $\lesssim 20$ MeV are supposed to be dominant for CCSN neutrinos. This is also supported from the aspect of weak interactions; the neutrino sphere of the high energy neutrinos is located at larger radii due to the large neutrino-matter cross sections. Since the matter temperature at the large radii is low ($\lesssim$ a few MeV), the neutrino emission above 50 MeV would be very weak. For these reasons, they have gained less attention from the community.

In this paper we study the energy dependent feature of neutrinos, focusing on the high energy region ($\gtrsim$ 50 MeV), where high energy neutrinos can be produced via a non-thermal process, {\it neutrino shock acceleration}. The possibility of the neutrino shock acceleration in CCSNe was first pointed out by \citet{Kazanas1981ICRC}, in which it was demonstrated by a Monte Carlo neutrino transport with an opacity of coherent scattering of $^{56}$Fe. \citet{Giovanoni1989ApJ} demonstrated the neutrino shock acceleration with more neutrino interactions, although the CCSN simulations and the input physics were based on old-fashioned models. It should be also mentioned that the neutrino shock acceleration is naturally included in CCSN simulations with fluid-velocity dependent spectral neutrino transport; indeed, we will show an evidence for neutrino acceleration seen in a CCSN simulation in Sec.~\ref{sec:theoryshockac}. However, there are  practical problems for  the accurate computation (see Sec.~\ref{sec:theoryshockac} for more details), which have prevented us from making the predictions for observable signatures of neutrino acceleration. On the other hand, as we shall show below, this process commonly occurs in all massive stellar collapse, indicating that this opens us a new possibility of detecting the high energy neutrinos by current- and future terrestrial detectors. In this paper, we reveal the conditions for which neutrino acceleration occurs, and also discuss the detectability of $\sim 100$ MeV neutrinos created by the shock acceleration, which have not been considered in previous studies. This paper is, hence, the first comprehensive study of the acceleration conditions, neutrino spectra, and observational consequences.

As we shall see in Sec.~\ref{sec:detect}, the detection of high energy neutrinos with $\gtrsim 100$ MeV will be an evidence that the neutrino shock acceleration occurs deep inside the CCSN core, and the neutrino spectrum would reflect the property of the shock wave. It should be also mentioned that the neutrino acceleration is strongly flavor-dependent, which is useful to prove the neutrino flavor conversion. As we shall describe in Sec.~\ref{sec:theoryshockac}, the high energy neutrinos can be only created for mu- and tau neutrinos (hereafter $\nu_{\mu}$ and $\nu_{\tau}$, respectively). Interestingly, the difference emerges even between $\nu_{\mu}$ and $\nu_{\tau}$ above the energy of $\sim 100$ MeV due to the appearance of charged-current reactions with muons. By virtue of the remarkable flavor dependent feature, the detection of high-energy neutrinos in terrestrial detectors through charged-current reaction channels (see Sec.~\ref{sec:detect} for more details) will be a smoking gun evidence that the neutrino goes through flavor conversions inside of the CCSN progenitor. Furthermore, charged-current reaction channels with $\nu_{\mu}$ and $\bar{\nu}_{\mu}$ would yield muons in detectors, which can be used to constrain neutrino mass hierarchy and the mixing angles, in principle. We discuss these observational consequence of the neutrino shock acceleration, although there remains work needed. More quantitative assessments require improvements in various elements, which will be also described in this paper.

This paper is organized as follows. In Sec.~\ref{sec:theoryshockac}, we first describe the mechanism of neutrino shock acceleration. We also present an evidence of the neutrino shock acceleration witnessed in previous CCSN simulations, which further strengthens our argument. We then demonstrate the neutrino shock acceleration by carrying out a post-processing neutrino transport simulations with a Monte Carlo transport method. The details of the input physics, setup and the results are summarized in Sec.~\ref{sec:MonteCarlo}. In Sec.~\ref{sec:detect}, we assess the detectability of high energy neutrinos with taking into account neutrino flavor conversions and detector configurations. Finally, we summarize our conclusion in Sec.~\ref{sec:sum}.

\section{Neutrino shock acceleration}\label{sec:theoryshockac}
\begin{figure}
\centering
\includegraphics[width=0.45\textwidth]{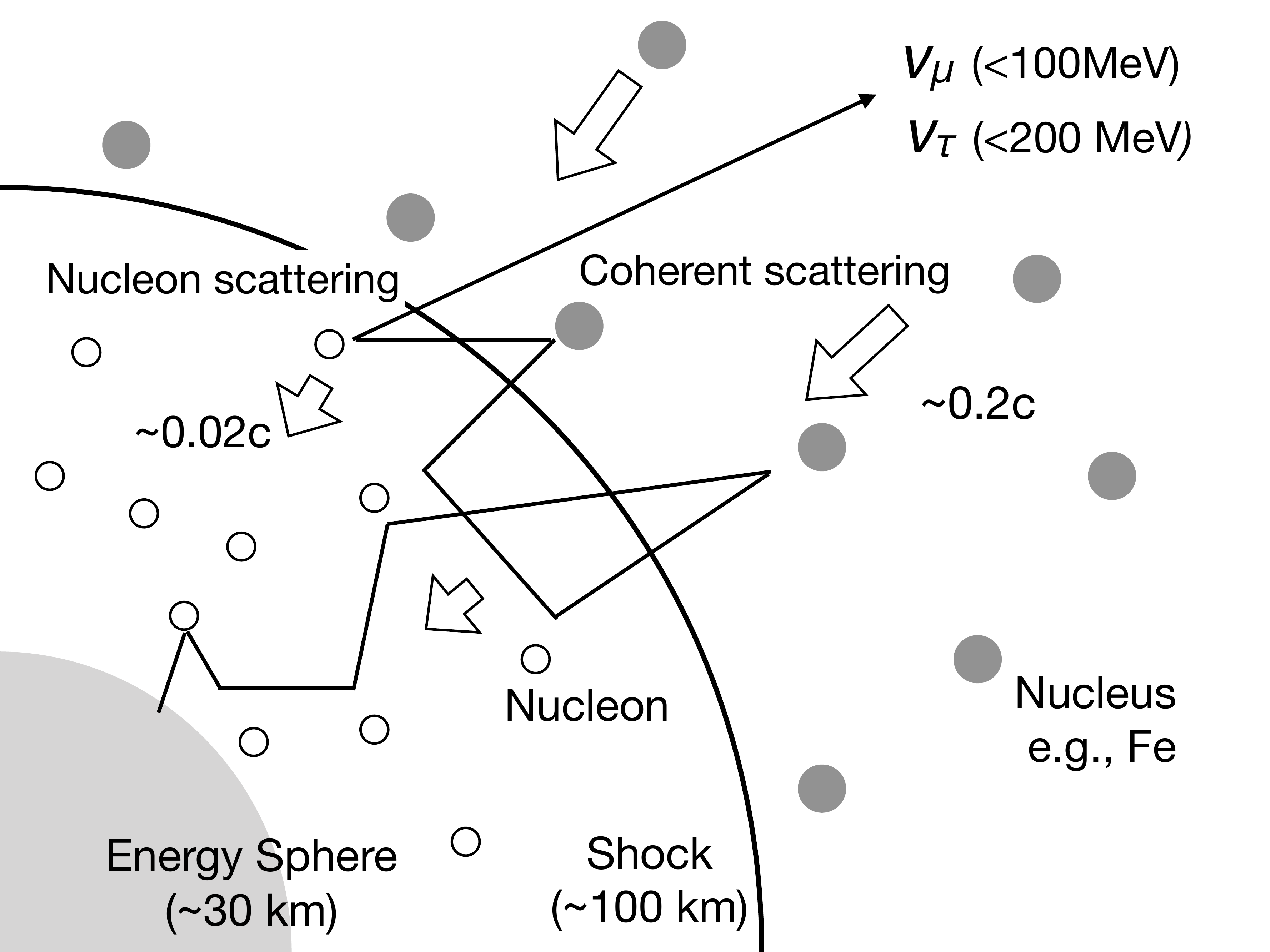}
\caption{Schematic picture of the neutrino shock acceleration in CCSN. Note that the shock acceleration occurs only for $\nu_{\mu}$, $\nu_{\tau}$ and their anti-partners (see the text for more details). Those neutrinos are emitted at the energy sphere (which is almost identical to the neutrino sphere and located at $\sim 30$ km) with a thermal spectrum. During the flight in the post-shock flows, they go through multiple-scatterings, which influences on the thermal spectrum, although the spectrum sustains the quasi-thermal feature \citep[see e.g.,][]{2013MNRAS.428.2443S,2019PTEP.2019h3E04S,2020ApJ...897...43K,2020PhRvD.102b3017W}. In the pre-shock region, the dominant opacity is coherent scatterings with heavy nuclei, on the other hand. Some neutrinos are back scattered by them and then cross the shock wave. In the post-shock flows, neutrinos have scatterings with nucleons again. Some fractions of neutrinos escape from the post-shock flows after repeating the same process (see Eq.~\ref{eq:tau} for the condition) during which neutrinos gain the energy from the shock wave and create the non-thermal spectrum. The reachable maximum energy is $\sim 100$ MeV and $\sim 200$ MeV for $\nu_{\mu}$, $\nu_{\tau}$, respectively. See the text for more details.}\label{fig:picture}
\end{figure}

Gravitational collapse of iron core abruptly ends when the inner core (homologously collapsing core) exceeds the nuclear density. The inner core rebounds due to the sharp rise of incompressibility, which forms a bounce shock wave at a mass coordinate of $\sim 0.5 M_{\sun}$. This region is opaque for all energies- and flavors of neutrinos, implying that neutrinos achieve thermal- and chemical equilibrium with matter. The shock wave propagates through supersonically infalling outer core and it transits to the semi-transparent region where neutrinos weakly couple with matter. This is the region where neutrinos experience the shock acceleration. Note that the shock acceleration occurs only if the neutrino absorption is negligible, i.e., heavy leptonic neutrinos are the leading players. To facilitate readers' understanding, we provide a schematic picture of the system in Fig.~\ref{fig:picture}.

The acceleration mechanism is essentially the same as the first-order Fermi-acceleration (diffusive shock acceleration), but neutrinos are scattered by nucleons, electrons and nuclei. Let us consider that a shock wave propagates in the scattering atmosphere at a radius of $r_{\rm sh}$ and the matter in the upstream (downstream) has a velocity of $v_{u}$ ($v_{d}$). The fractional change of the neutrino's energy, $E$, at each shock crossing is roughly 
\begin{eqnarray}
\left\langle \frac{\Delta E}{E} \right\rangle \approx \frac{\left|v_u-v_d\right|}{c},\label{eq:gain}
\end{eqnarray}
where $c$ is the speed of light and the symbol of $\left\langle \right\rangle$ denotes the average. The acceleration stops when the energy loss on each scattering becomes comparable to the gained energy (Eq.~\ref{eq:gain}) or the accelerated particle is absorbed by matter. For electron-type neutrinos ($\nu_{e}$) and their anti-partners ($\bar{\nu}_{e}$), no shock accelerations occur in practice, since their shock acceleration is hampered by the reactions of $\nu_{e} + n \rightarrow e^{-} + p$ and $\bar{\nu}_{e} + p \rightarrow e^{+} + n$. The cross section for these reactions increases with energy, implying that the accelerated neutrinos for $\nu_{e}$ and $\bar{\nu}_{e}$ would be immediately absorbed.

\begin{figure}
\centering
\includegraphics[width=0.5\textwidth]{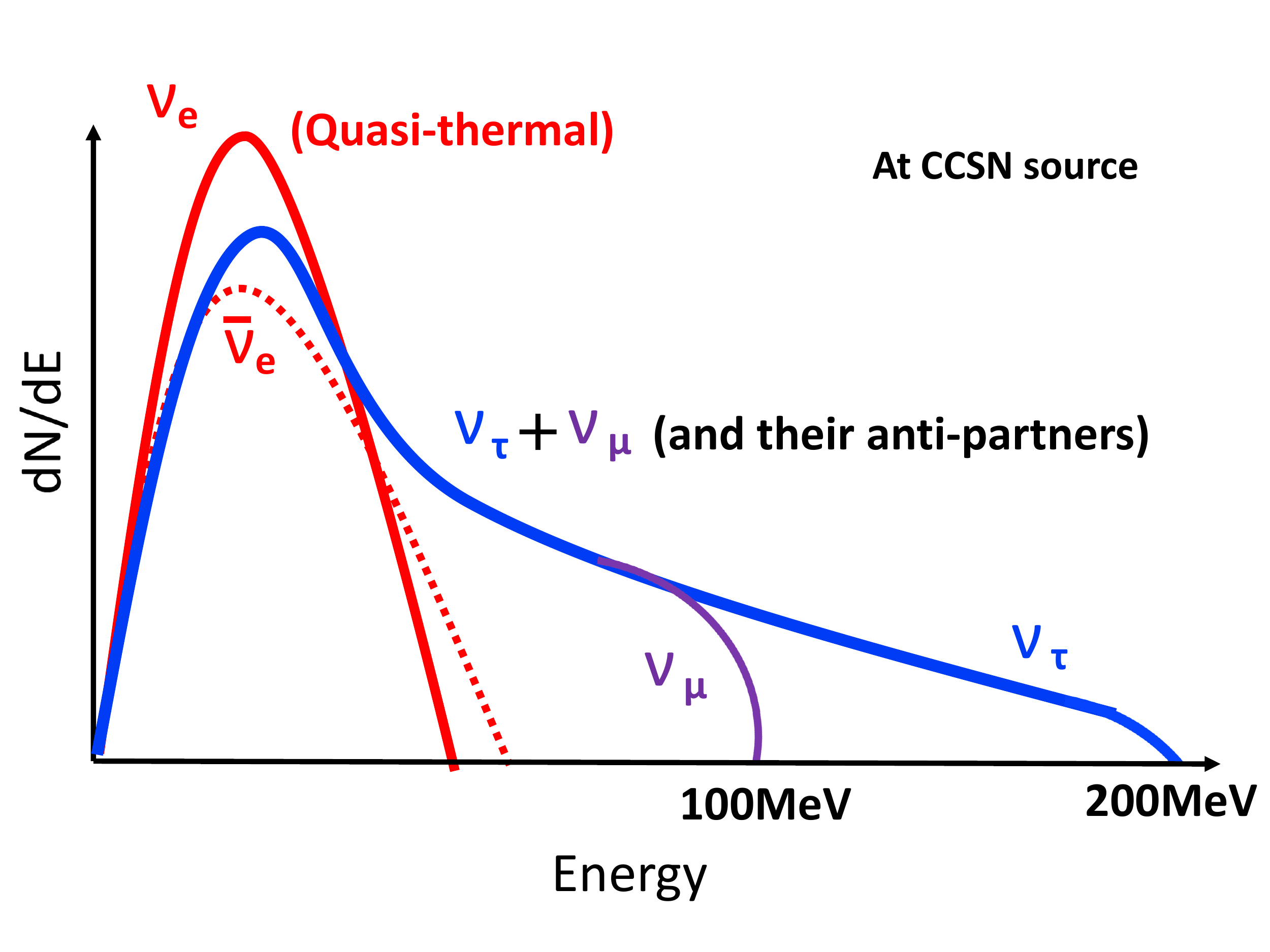}
\caption{Schematic picture of expected spectra for all flavors of neutrinos (at a CCSN source) when the neutrino shock acceleration occurs.}\label{SchematicPic}
\end{figure}

For $\nu_{\mu}$ and $\nu_{\tau}$, on the other hand, the situation is very different from that of the electron-type (see also Fig.~\ref{fig:picture}), since their charged current reactions are absent at least up to $\sim 100$ MeV neutrinos\footnote{Strictly speaking, weak processes such as neutrino pair annihilations and the inverse process of nucleon-nucleon  bremsstrahlung work similar as neutrino absorptions by matter. However, those reactions are negligible for the considered situation.}. Let us first estimate the upper limit of neutrino energy ($E_{\rm max}$) without absorption processes. This is mainly determined by the balance between the energy loss of scatterings and the energy gain by the shock acceleration (Eq.~\ref{eq:gain}) , i.e.,
\begin{eqnarray}
\frac{\left|v_u-v_d\right|}{c} \sim \frac{E_{\rm max}}{M},\label{eq:loss}
\end{eqnarray}
where $M$ is the mass of the scattering particle, i.e., nucleons and nucleus in the downstream and upstream respectively. Since nucleons have the lighter mass, $1\,{\rm GeV}$, the shock acceleration is limited by the nucleon recoil effect in the downstream\footnote{But see Appendix for a discussion regarding the effect of light nuclei.}. Since the fluid-velocity difference between up- and down stream at the shock wave is $\approx 0.2c$, we obtain $E_{\rm max} \sim 200$ MeV\footnote{We note that the nucleus-neutrino inelastic scattering in the upstream and the electron scattering in the downstream should be taken into account for more quantitative arguments; indeed they reduce the efficiency of the shock acceleration. However, this effect does not change significantly our discussion. We will discuss these effect in Sec.~\ref{sec:MonteCarlo} for more details.}. This is smaller than the mass of tau ($\sim 1$ GeV), indicating that $\nu_{\tau}$ can be in principle accelerated up to $ \sim 200$ MeV.    On the other hand, the $\nu_{\mu}$ acceleration stops before reaching $\sim 200$ MeV, since the $\nu_{\mu}$ absorption via charged-current reactions, e.g., $\nu_{\mu}+n\rightarrow p+\mu^-$ and $\bar{\nu}_{\mu}+p\rightarrow n+\mu^+$, emerge above $\sim 100$ MeV. $\nu_{\mu}$ are mainly absorbed by nucleons and nuclei in the down- and upstream, respectively, meanwhile muons produced in these processes decay immediately and emit $\nu_{\mu}$ at the lower energy ($\ll 100$\,MeV). Consequently, the spectrum of $\nu_{\mu}$ is expected to have a sharp cut off around $100$ MeV. This fact suggests that the energy spectrum is remarkably different between $\nu_{\mu}$ and $\nu_{\tau}$ in the energy range of $100$ MeV $ \lesssim E \lesssim 200$ MeV, which provides a precious environment to study the neutrino oscillation in CCSN (see Sec.\ref{sec:detect} for more details). Note that the numbers of shock crossings for which a seed neutrino with an energy of $50\,{\rm MeV}$ is accelerated to $100$ and $200\,{\rm MeV}$ are roughly $5$ and $10$, respectively.  As a summary, we draw a schematic picture of the expected feature of neutrino number spectrum around the shock wave in Fig.~\ref{SchematicPic}.

To assess the detectability of the high energy neutrinos created by the neutrino shock acceleration, it is necessary to consider the acceleration time scale and the escape probability of the neutrinos from the system.
To estimate the time scale, we first take a look at the optical depth ($\tau_s$),
\begin{eqnarray}
\tau_s(t,E) & = & \int_{r_{\rm sh}(t)}^{\infty}dr \sigma(E)n(t,r),\\
& \approx & 1 \left(\frac{r_{\rm sh}}{100\,{\rm km}}\right)^{-3.4}\left(\frac{E}{50\,{\rm MeV}}\right)^{\alpha}, \label{eq:tau_s_edepe}
\end{eqnarray}
where $\sigma$ is the cross section of neutrino-nucleus scattering and we assume that neutrinos are scattered by iron via neutral-current reactions. 
Here we use the number density of nucleons in post shock region, $\mbox{$r\leq r_{\rm sh}$}$, and nuclei in pre-shock region, $\mbox{$r> r_{\rm sh}$}$, described by Eq. (\ref{eq:n}) in Appendix. The time evolution of the shock radius is set as $r_{\rm sh}=100\,{\rm km}(t/20\,{\rm ms})^{0.4}$, which is a reasonable approximation to reproduce that demonstrated in our CCSN simulations \citep{2019ApJS..240...38N}.
Equation~\ref{eq:tau_s_edepe} suggests that neutrinos with $\gtrsim 50$ MeV are the injected neutrinos around $20\,{\rm ms}$ and the acceleration time scale can be roughly estimated as $\lesssim N_s r_{\rm sh}/c \sim 3(N_s/10)$ ms, where $N_s$ is the number of shock crossings\footnote{It should be noted that this is a conservative estimate. It is attributed to the fact that the scattering cross section monotonically increases with neutrino energy, indicating that the neutrino flight distance between scatterings would be smaller with increasing energy in reality. In the above estimation, however, the flight distance is set as a constant ($r_{\rm sh}$), which provides the upper limit of acceleration time scale. }. It is an order of magnitude shorter than the post-bounce time, indicating that the system has long enough time to generate non-thermal neutrinos.

\begin{figure}
\centering
\includegraphics[width=0.45\textwidth]{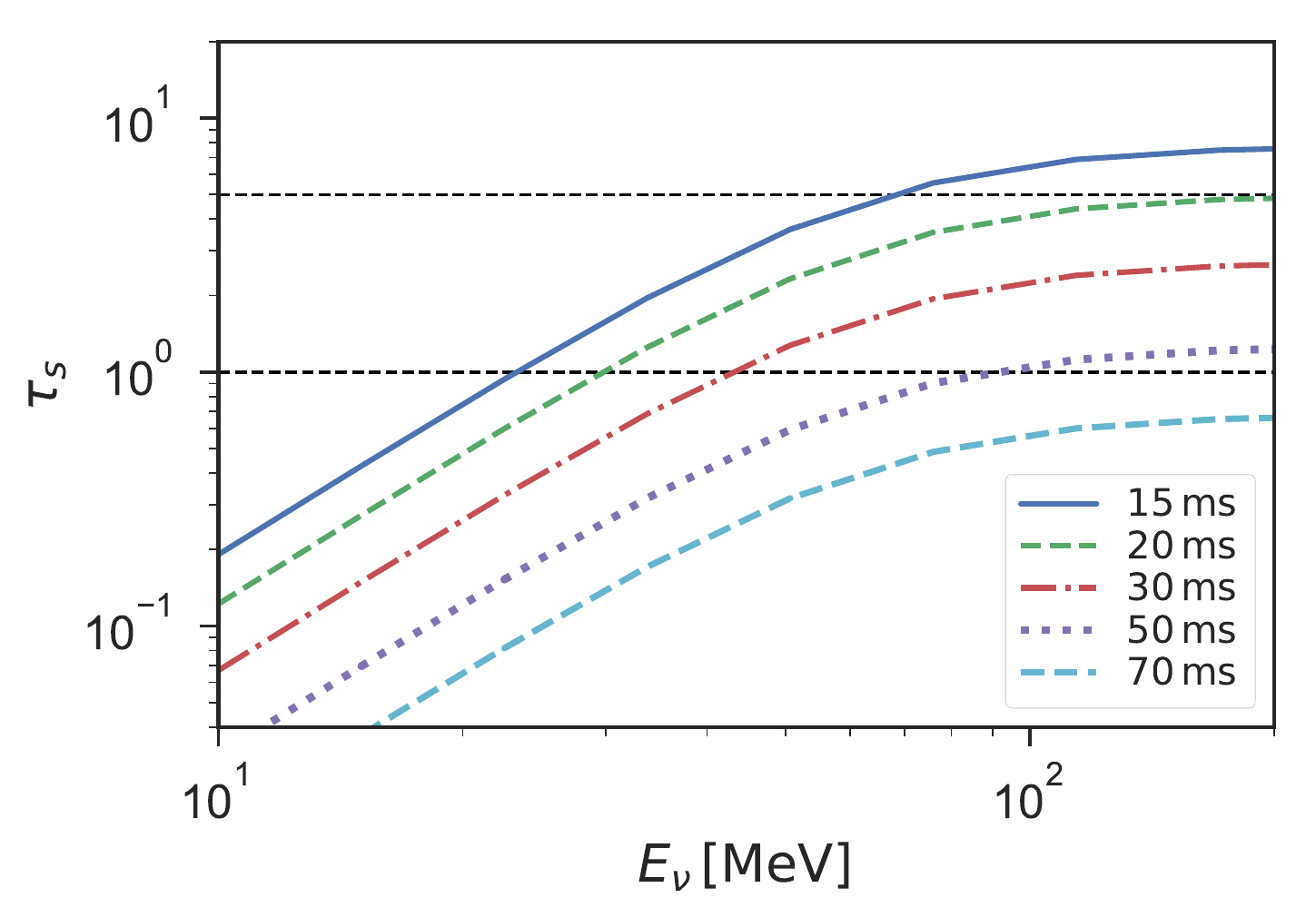}
\caption{Neutrino optical depth to coherent scatterings with heavy nuclei at the shock.}\label{fig:tau}
\end{figure}

On the other hand, the accelerated neutrinos tend to be trapped by fluid and be advected to the downstream region, since the cross section of neutrino-matter interaction increases with neutrino energy. Therefore, the observable neutrinos need to satisfy the following conditions: (i) the accelerated neutrinos can escape within a dynamical time scale and (ii) the scattering probability in the upstream is larger than unity. These conditions correspond to 
\begin{eqnarray}
1\lesssim \tau_s(t,E) \lesssim \frac{c}{v_u}.\label{eq:tau}
\end{eqnarray}
Fig.~\ref{fig:tau} shows the time evolution of $\tau_s$ with respect to neutrino energy. Note that the cross section of scatterings with nuclei increases with energy\footnote{The rate of increase becomes mild in the energy of $\gtrsim 50$ MeV, which is attributed to the fact that the neutrino scattering with nuclei looses the coherency with increasing neutrino energy.} and the power law index, $\alpha$, is $0.5\lesssim \alpha \lesssim 2$ in the relevant energy range. Roughly speaking, high energy neutrinos ($> 50$ MeV) are expected to emerge when the shock radius is located at $\sim 100$ km. However, the evaluation of the emergent spectrum is not trivial since (1) both the efficiency of neutrino shock acceleration and the escape probability strongly depend on the neutrino energy; (2) the two effects are competing each other; (3) the system is not steady, implying that the time-dependent matter background should be taken into account. We hence address the issue by carrying out post-processing neutrino transport simulations by employing a Monte Carlo transport method. See the next section for more details.

\begin{figure}
\centering
\includegraphics[width=0.45\textwidth]{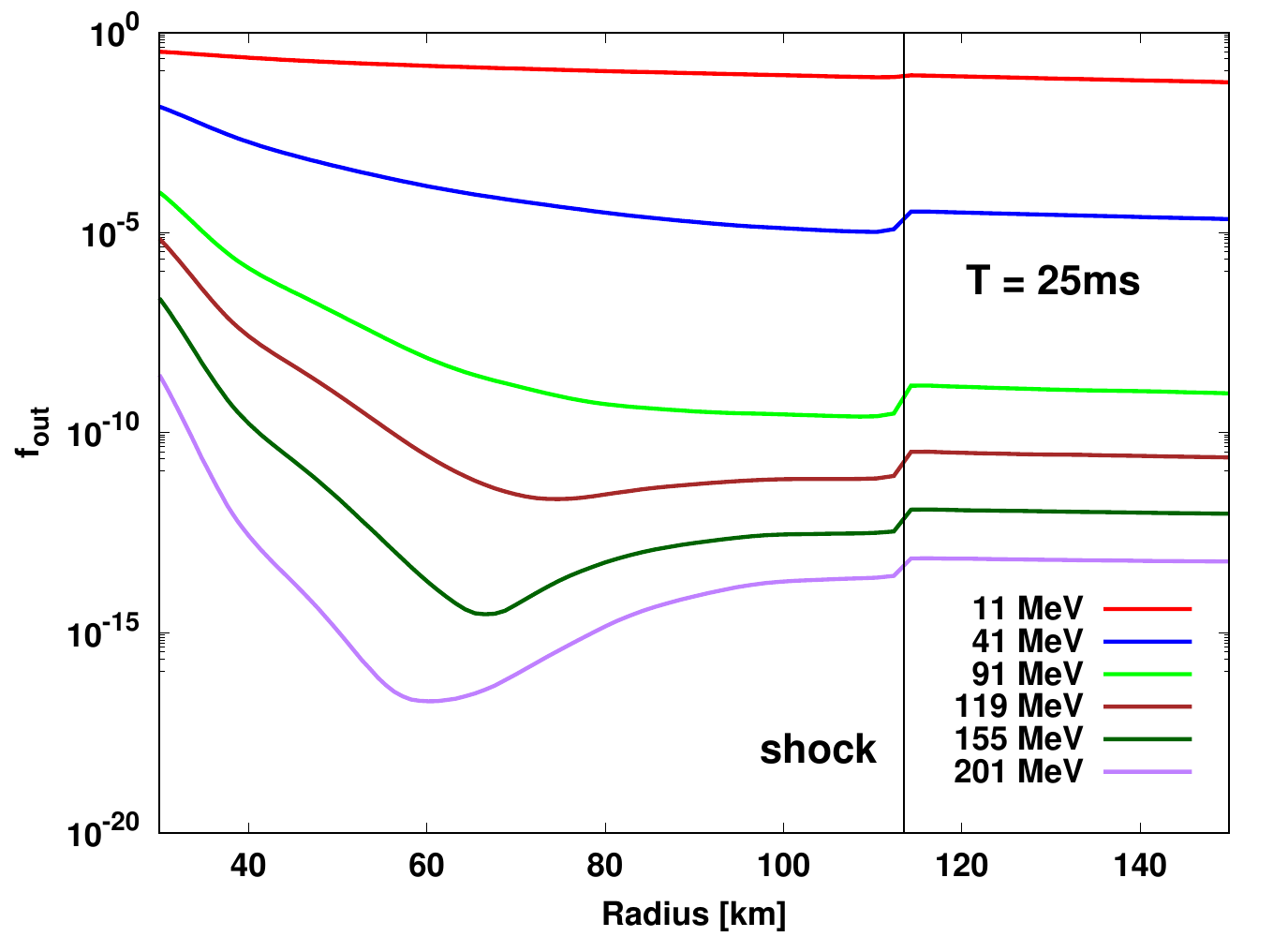}
\caption{Radial profiles of neutrino distribution function for out-going heavy leptonic neutrinos obtained in a CCSN simulation with full Boltzmann neutrino transport \citep{2019ApJS..240...38N}. The color represents the neutrino energy (measured in the fluid rest frame), and the thin-vertical line at $\sim 113$ km denotes the shock radius. The non-monotonic radial profile is observed in the post shock region for high energy neutrinos, which corresponds to an evidence of the occurrence of neutrino shock acceleration. See the text for more details. }\label{graph_fout_nux_radialdistri_25ms_forpaper}
\end{figure}

We make a few remarks here. Regarding neutrino transport simulations, it is, in principle, possible to demonstrate the neutrino shock acceleration by neutrino-radiation hydrodynamics simulations of CCSN. As a matter of fact, we have witnessed the neutrino shock acceleration in our CCSN simulations with full Boltzmann neutrino transport \citep{2018ApJ...854..136N,2019ApJS..240...38N,2019ApJ...880L..28N}. In these simulations, the special relativistic effects (fluid-velocity dependence) in neutrino transport are incorporated consistently \citep[see also][]{2014ApJS..214...16N}, indicating that the neutrino shock acceleration naturally occurs. In Fig.~\ref{graph_fout_nux_radialdistri_25ms_forpaper}, we display an example: radial profiles of out-going neutrino distribution function ($f_{\rm out}$) for some selected energies (measured on the fluid-rest frame) of heavy leptonic neutrinos. The corresponding model is a 15 $M_{\sun}$ progenitor in \citet{2019ApJS..240...38N} and we select a time snapshot of $25$ ms after bounce. As shown in the figure, $f_{\rm out}$ for low energy neutrinos monotonically decreases with radius, which is mainly attributed to the fact that some fractions of neutrinos go through out-scatterings by matter. On the other hand, $f_{\rm out}$ with $E > 91$ MeV has a non-monotonic radial profile; it decreases with radius up to $r \sim 70$ km but the opposite trend emerges in between $r \sim 70$ km and the shock radius (it increases with radius). The emergence of the positive radial gradient of $f_{\rm out}$ corresponds to the sign of the neutrino shock acceleration. Once the shock wave enters the scattering atmosphere, the neutrino shock acceleration kicks in. The number density of high energy neutrinos increases with time as long as the condition of $1\lesssim \tau_s(t,E)$ is satisfied (see also Eq.~\ref{eq:tau}), implying that the resultant radial gradient of $f_{\rm out}$ becomes positive\footnote{Strictly speaking, electron scatterings and other weak processes such as pair and brems also account for the positive gradient of $f_{\rm out}$. Indeed, heavy leptonic neutrinos are eventually thermalized deep inside the post-shock region by these processes.} (in other words, the time evolution of $f_{\rm out}$ is imprinted in the radial profile).

Aside from ours, the non-thermal neutrinos have been also observed in other CCSN simulations \citep[see, e.g.,][]{2007PhR...442...38J,2015ApJ...804...75N}. It should be noted, however, that the obtained neutrino spectrum in these CCSN simulations is not accurate for several reasons. The most serious problem is the neutrino matter interactions incorporated in these simulations. First, we did not include charged-current reactions with $\nu_{\mu}$, implying that its spectrum is no longer accurate in the energy of $> 100$ MeV\footnote{Note that the heavy leptonic neutrinos are treated collectively in most CCSN simulations, which is also another deficit to discuss the neutrino shock acceleration.}. It should be also mentioned that nucleon recoils are not incorporated in these simulations. As described in \citet{2020ApJ...897...43K}, there is a practical problem of handling the recoils by finite volume methods of neutrino transport (like our Boltzmann code), since unfeasibly high energy-resolutions in neutrino transport are required to handle the small energy exchange between neutrinos and matter accurately. Note that, although some approximate treatments for the non-isoenergetic scatterings have been adopted \citep[see, e.g.,][]{2002PhRvD..65d3001H} in some CCSN simulations, the assumptions are no longer valid for neutrinos $\gtrsim 50$ MeV, which is a crucial problem for quantitative arguments of neutrino shock acceleration. For these reasons, we take a particle-based approach (Monte Carlo transport) that is more appropriate than any finite-volume methods to handle the neutrino matter interactions in the considered situation.

One may wonder that stable muon creations, which have been recently discussed in the literature \citep[see, e.g.,][]{2017PhRvL.119x2702B,2020arXiv200813628F,2020PhRvD.102b3037G}, suppress the neutrino shock acceleration. We note, however, that it does not directly affect the neutrino shock acceleration, since the region where the shock acceleration occurs is much low density and low temperature environment compared to those with the stable muon creations. This assures that the neglecting these weak processes is reasonable. On the other hand, muon creations may affect indirectly through the change of the thermal component of the spectrum; indeed, the chemical potential of $\nu_{\mu}$ (and $\bar{\nu}_{\mu}$) is not zero in the case. This also potentially distinguish the spectrum of $\nu_{\mu}$, $\nu_{\tau}$, and their anti-partners. It would be interesting to incorporate those effects in our study, although it is beyond the scope of this paper.

As a final but important remark, we speculate that the neutrino shock acceleration potentially occurs in failed CCSN. In this case, the shock wave is confined to the region of ($\sim 100$ km) due to high ram pressure of mass accretion rate; consequently the PNS will eventually collapse onto a BH. There may be a certain time window that the condition of Eq.~\ref{eq:tau} is satisfied for $\gtrsim 50$ MeV neutrinos. It should be mentioned, however, that there are three major differences from the early phase. The first is that the system is quasi-steady, i.e., we can neglect the time evolution of fluid background. Second, the thermal neutrinos have higher temperature and may be more abundant than those in the early phase, since the PNS is more compact and larger accretion energy would be released in the case. They increase the number of injected (seed) neutrinos at $\sim 50$ MeV. Third, the shock wave may be highly deformed by multi-dimensional fluid instabilities such as Standing Accretion Shock Instability (SASI). This indicates that the accretion flow passes through the shock front obliquely, which reduces the conversion efficiency from kinetic energy to thermal one. In another word, the fluid-velocity difference between up- and down stream becomes smaller, which results in reducing the efficiency of neutrino shock acceleration. In this case, the reachable maximum energy of $\nu_{\tau}$ would be smaller than $\sim 200$ MeV (see Eq.~\ref{eq:loss}). It should be noted, however, that the morphology of shock wave is stochastically changed with time, and the amplitude of fluctuations depends on the vigor of fluid instabilities, which also depends on the time and place. To address these issues, we need sophisticated multi-D models of CCSN and then carry out the systematic study of the impact on the shock acceleration. The detailed study is much beyond the scope of this paper, hence, postponed to future work. With keeping in mind the uncertainty, we discuss the neutrino shock acceleration and the observational consequence under spherically symmetric and steady-state approximation in Sec.~\ref{subsec:deteBHcase}.

\section{Monte Carlo simulation}\label{sec:MonteCarlo}
To calculate the spectra of the emergent mu and tau neutrinos, we run Monte Carlo simulations for the neutrino transport in CCSN.  The simulation takes into account for the energy loss due to the recoil effect and inelastic scattering as well as thermalization.
We include the following processes:\\
(i) nucleon scattering \citep{Beacom2002}
\begin{eqnarray}
n + \nu_X \rightarrow n+\nu_X,
\end{eqnarray}
(ii) coherent scattering \citep{Papoulias2018}
\begin{eqnarray}
A + \nu_X \rightarrow A+\nu_X,
\end{eqnarray}
(iii) nucleus inelastic scattering \citep{Dapo2012}
\begin{eqnarray}
A + \nu_X \rightarrow A^{*}+\nu_X,
\end{eqnarray}
(iv) electron scattering \citep{Tubbs1975,Bowers1982A}
\begin{eqnarray}
e^- + \nu_X \rightarrow e^-+\nu_X,
\end{eqnarray}
(v) quasi-nucleon scattering \citep{Formaggio2012RvMP}
\begin{eqnarray}
n + \nu_{\mu} \rightarrow p+\mu^{-},
\end{eqnarray}
(vi) quasi-nucleus scattering
\begin{eqnarray}
(A,Z) + \nu_{\mu} \rightarrow (A,Z+1)+\mu^{-}.
\end{eqnarray}
 We use the total cross section of nucleus inelastic scattering given by \cite{Dapo2012} up to $100\,{\rm MeV}$ and extrapolate it as $\propto E_{\nu}^2$. We assume that
 neutrinos lose  $10\,{\rm MeV}$ in each nucleus-inelastic scattering and are emitted isotropically in the nucleus restframe.
 Quasi-nucleon and quasi-nucleus scatterings are effective only at high energies, $E_{\nu}>m_{\mu}$. The neutrino energy after a quasi scattering is assumed to be one third of the incident neutrino energy. The cross section of quasi-nuclear scattering is simply assumed to be $A_{\rm nuc}\sigma_q/2$, where $\sigma_q$ is the cross section of quasi-nucleon scattering and $A_{\rm nuc}$ is the mass number of the nucleus.
 Note that we neglect other processes such pair annihilation and bremsstrahlung that are much weaker than the above processes in the region considered here.

In the early post-bounce phase, $t\sim 10\,{\rm ms}$, the fluid background evolves with a comparable or short time scale to that of the neutrino shock acceleration, implying that the matter evolution should be also taken into account appropriately. Our Monte Carlo neutrino transport is solved with dynamical fluid background; the evolution of the matter quantities such as density, velocity and temperature are described analytic functions that approximate the results of a spherically symmetric simulation with Boltzmann neutrino transport \citep{2019ApJS..240...38N}. The details of these functions are shown in Appendix. For the late post bounce phase in failed CCSN, on the other hand, we impose a steady state approximation in fluid background, which can be characterized by the mass accretion rate (see below). For simplicity, the downstream and upstream are assumed to be composed of nucleons and $^{56}$Fe, respectively. Note that we neglect the blocking effect because we focus on the high energy tail of the neutrino spectra, where $f\ll 1$.

We inject in total $4\cdot 10^6$ neutrino particles at $50\,{\rm km}$ with a thermal spectral of zero chemical potential with $T=6\,{\rm MeV}$. The injection time of each particle is chosen from a certain time window after the bounce with a uniform distribution, corresponding to a constant neutrino luminosity. We stop following each Monte Carlo particle when it reaches either the inner boundary $r=30\,{\rm km}$ or the outer boundary $r=700$ km. The particles reach the outer radius are considered to  freely escape afterward. 

\begin{figure}
\centering
\includegraphics[width=0.45\textwidth]{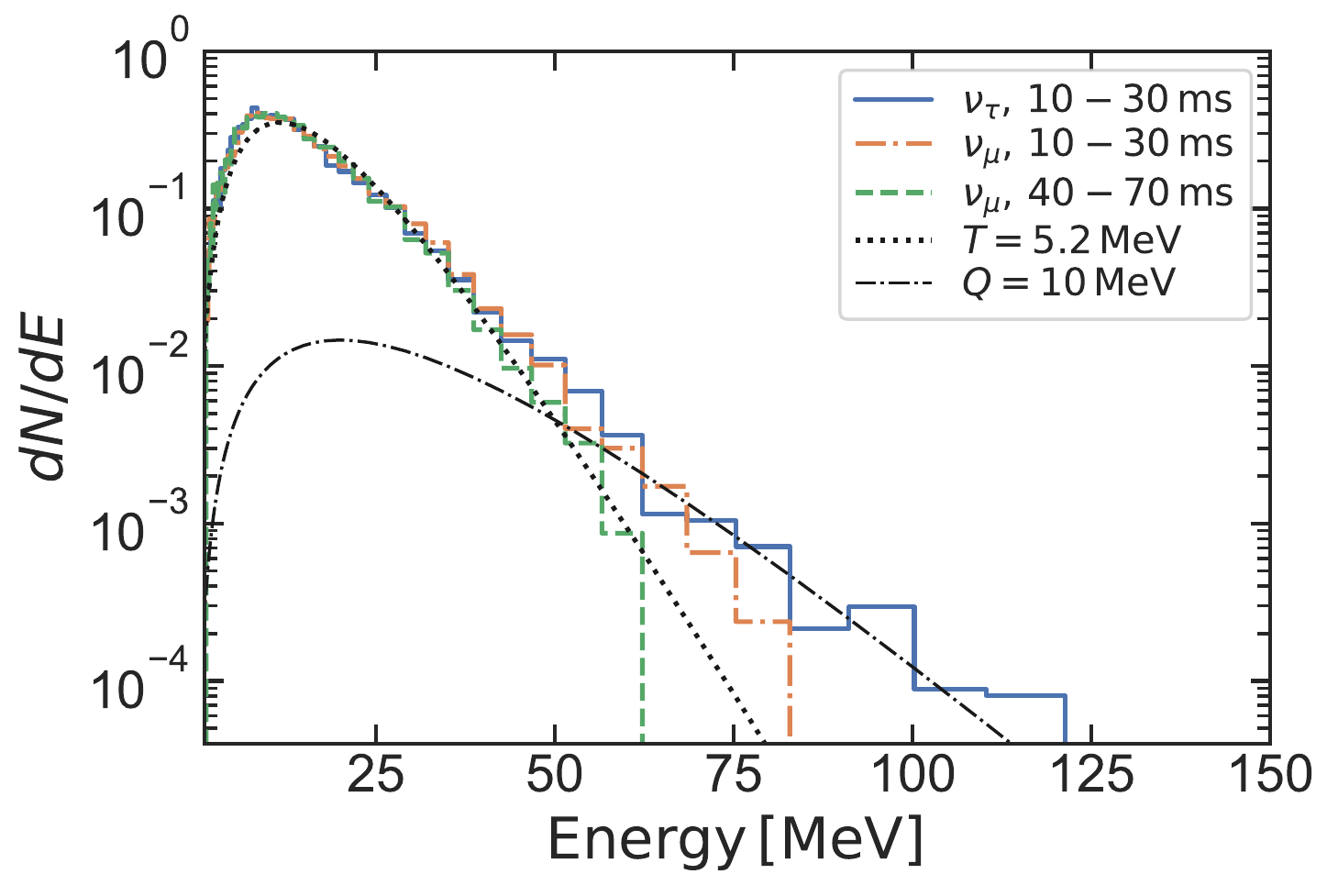}
\caption{Emergent spectra of $\nu_{\mu}$ and $\nu_{\tau}$ at $10$--$30,{\rm ms}$ after the bounce, obtained by our Monte Carlo simulations. For comparison, we display a $\nu_{\mu}$ spectrum at $40$--$70,{\rm ms}$ after the bounce, a thermal spectrum of zero chemical potential with $T=5.2\,{\rm MeV}$, and a function given by Eq. \ref{eq:spectfit_nt}. with $Q=10\,{\rm MeV}$.}\label{fig:spec1}
\end{figure}

\begin{figure}
\centering
\includegraphics[width=0.45\textwidth]{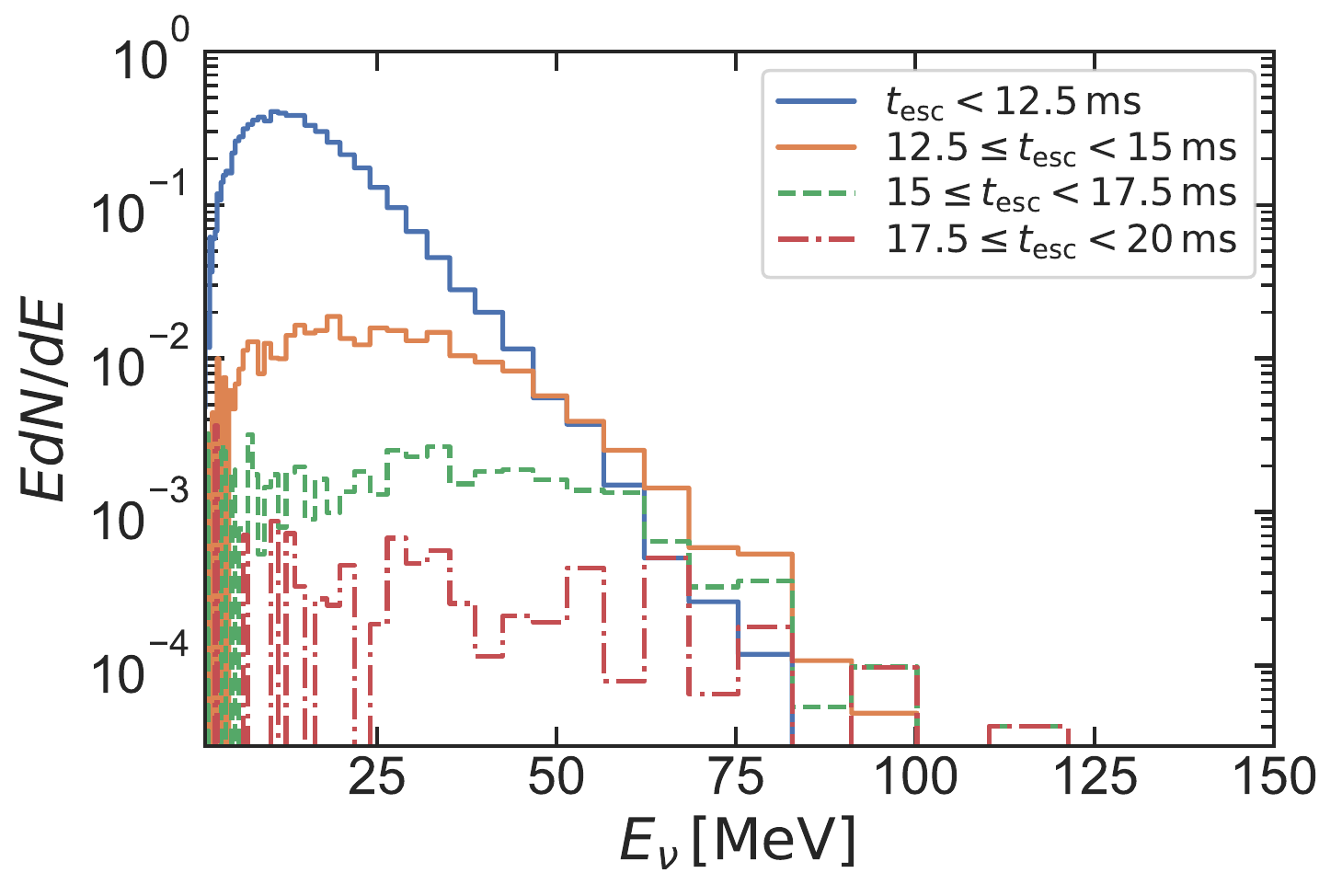}
\caption{Emergent spectra of $\nu_{\tau}$ at different arrival times 
for the early post bounce phase of a CCSN. Here we inject neutrinos at $t=10\,{\rm ms}$ after the bounce and observe the emergent neutrino spectra at a radius of $150\,{\rm km}$. }\label{fig:tacc}
\end{figure}

Fig.~\ref{fig:spec1} shows the $\nu_{\mu}$ and $\nu_{\tau}$ spectra $10$--$30\,{\rm ms}$ after the bounce. For comparison, we also show a thermal spectrum of zero chemical potential with $5.2\,{\rm MeV}$, which describes the numerical spectra around the peak. Both the $\nu_{\mu}$ and $\nu_{\tau}$ spectra significantly exceed the thermal spectral above $\sim 50\,{\rm MeV}$.
The $\nu_{\mu}$ spectrum has a sharp cut off at $\sim 100\,{\rm MeV}$ because the cross section of muon creation sharply rises\footnote{We observe the cut off in the $\nu_{\mu}$ spectrum in Fig. \ref{fig:spec1} slightly less than the muon restmass because $\nu_{\mu}$ in the lab frame is blue shifted in the fluid restframe in the upstream.}. The magnitudes of the non-thermal tail relative to the flux expected from the the thermal distribution around $80\,{\rm MeV}$ are approximately $(8,8,13,5,3)$ at $t=(15,20,25,30,35)\,{\rm ms}$ post bounce. As we discussed in the previous section, the effect of the neutrino acceleration to the observed spectra is the most significant around $25\,{\rm ms}$ post bounce corresponding to $\tau\sim c/v_u$ at $\sim 100\,{\rm MeV}$. Fig.~\ref{fig:tacc} shows the emergent $\nu_{\tau}$ spectra at different escape times, $t_{\rm esc}$, at $150\,{\rm km}$, where we inject all neutrinos at $t=10\,{\rm ms}$. The low-energy neutrinos escape quickly, implying that these neutrinos do not experience any scatterings. On the contrary,  the neutrinos in the high-energy tail escapes at $\gtrsim  2.5\,{\rm ms}$, which agrees with our estimation discussed in the previous section.

\begin{figure}
\centering
\includegraphics[width=0.45\textwidth]{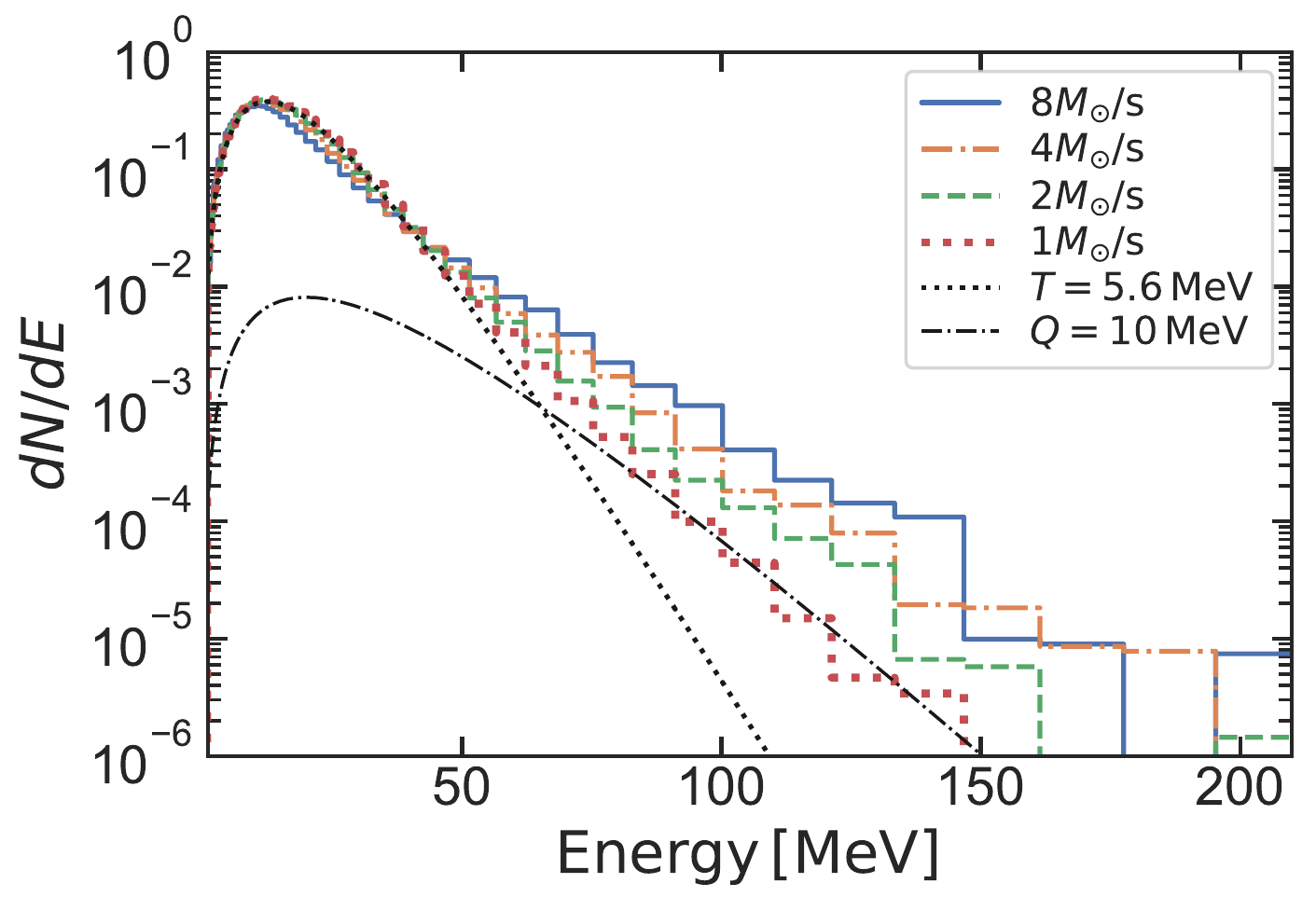}
\caption{Same as Fig.~\ref{fig:spec1} but for the late post bounce phase in failed CCSN with different mass accretion rates $\dot{M}$.}\label{fig:spec2}
\end{figure}

We also study the case of failed CCSN, in which we assume that
a shock wave stays at $80\,{\rm km}$. The velocity of the upstream at $80\,{\rm km}$ is set to be $0.2c$ and the density at the shock front is 
characterized by the mass accretion rate, $\dot{M}$.
The radial profile of the former is assumed to be proportional to $r^{-0.5}$ and the latter is determined from $\dot{M}={\rm constant}$. Unlike the CCSN model, the density and velocity profiles of the failed CCSN model are assumed to be stationary. 
Fig.~\ref{fig:spec2} shows the $\nu_{\tau}$ spectra in the case of failed CCSN. The emerging spectrum depends on the mass accretion rate. 
Roughly speaking, the neutrino shock acceleration occurs when $(\dot{M}/1M_{\odot}/s)(80\,{\rm km}/r_{\rm sh})\gtrsim 1$ is satisfied.

\begin{figure}
\centering
\includegraphics[width=0.45\textwidth]{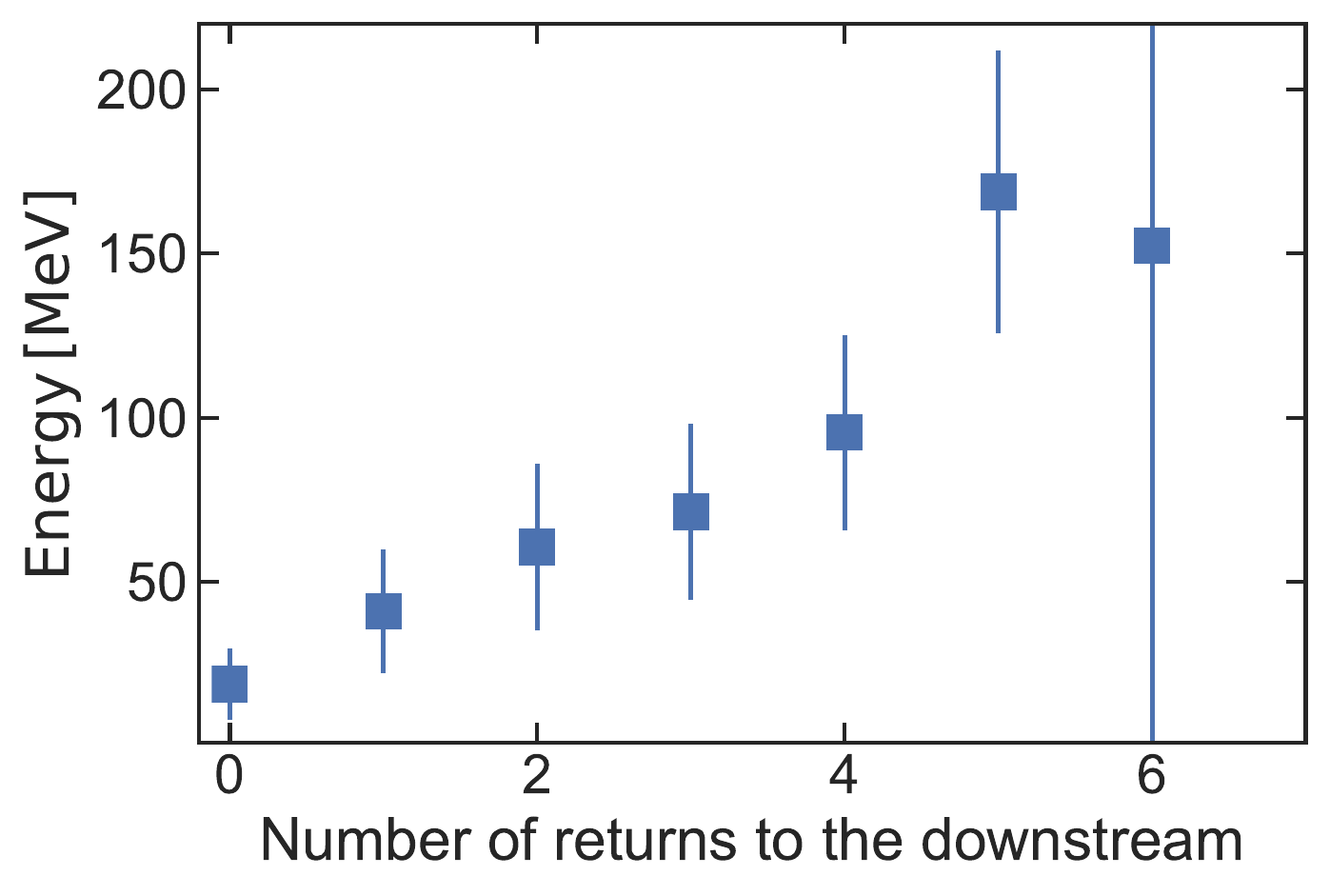}
\caption{Energy of emergent neutrinos as a function of the number of returns to the downstream in the case of failed CCSN with $\dot{M}=2M_{\odot}/s$. The mean neutrino energies and mean absolute deviations are shown as squares and bars. }\label{fig:shock}
\end{figure}

We note that neutrinos gain the energy due to the converging flow in the upstream, which potentially contribute to the non-thermal spectra \citep{1981MNRAS.194.1033B,2013MNRAS.428.2443S}. This mechanism works, however, only in the case that the shock is located deep inside the scattering atmosphere, $\ll 100\,{\rm km}$. On the other hand, most of accelerated neutrinos advect inwards in such a case, indicating that they give less impact on observable neutrinos. To assess this argument, we analyze the relation between number of returns to the downstream and neutrino energies in the case of the failed CCSN with $2M_{\odot}/s$; the result is shown in Fig.~\ref{fig:shock}. The positive correlation between these two quantities can be clearly seen in the figure. We, thus, conclude that the (observable) non-thermal neutrinos which we found in our Monte Carlo simulations are primary created by shock acceleration, which is consistent with our analytic discussion made in Sec.~\ref{sec:theoryshockac}.

\section{Detectability}\label{sec:detect}
We assess the detectability of the high energy neutrinos in some representative terrestrial neutrino detectors. We first describe basic assumptions for computing the event counts on each detector in Sec.~\ref{subsec:basicasump}. We then present the results for the early post bounce phase in Sec.~\ref{subsec:deteearlyresults}. The similar estimation but for the later phase in the case with failed CCSN is presented in Sec.~\ref{subsec:deteBHcase}. Finally, we discuss a possibility of muon productions in these detectors in Sec.~\ref{subsec:muondetec}.

\subsection{Basic assumptions}\label{subsec:basicasump}
\subsubsection{Analytic expression of neutrino spectrum}\label{subsubsec:anaexpress}
As described in Sec.~\ref{sec:theoryshockac}, the neutrino shock acceleration does not occur for $\nu_e$ and $\bar{\nu}_e$, indicating that their spectrum remains quasi-thermal feature (see also Fig.~\ref{SchematicPic}). We also note that their average energy tends to be smaller than that of the heavy leptonic neutrinos, which indicates that the exponential decline of the thermal spectrum in the high energy region is steeper than that of heavy leptonic neutrinos. For these reasons, $\nu_e$ and $\bar{\nu}_e$ in the high energy region ($> 50$ MeV) would be subdominant; hence we ignore their contributions in this estimation.

In this study, we do not directly use the raw data of the emergent neutrino spectrum computed by Monte Carlo simulations but rather fit them by an analytic formula to facilitate intuitive understanding of the event count. In the expression, the number spectrum of neutrinos at the CCSN source ($dN/dE$ [MeV$^{-1}$]) is assumed to be the sum of thermal- and non-thermal component,
\begin{eqnarray}
\frac{dN}{dE} = \left( \frac{dN}{dE} \right)_{\rm th} + \left( \frac{dN}{dE} \right)_{\rm nt}
,\label{eq:spectfit_tot}
\end{eqnarray}
where
\begin{eqnarray}
\left( \frac{dN}{dE} \right)_{\rm th} &=& A \hspace{1mm}  \frac{E^2}{1+{\rm exp}( E/T  )},\label{eq:spectfit_th} \\
\left( \frac{dN}{dE} \right)_{\rm nt} &=& B \hspace{1mm}  E^2  {\rm exp}( -E/Q  ).\label{eq:spectfit_nt}
\end{eqnarray}
In these expressions, $E$ and $T$ denote the neutrino energy and temperature, respectively. $Q$ represents the exponent of the high energy tail of the non-thermal component. The coefficients of $A$ and $B$ determine the scale of neutrino luminosity on each component. In this study, we assume that the number spectrum of $\nu_{\mu}$ and $\nu_{\tau}$ (and their anti-partners) is identical each other up to the energy of muon mass ($106\,{\rm MeV}$). In the above energy, the spectrum of $\nu_{\tau}$ extends up to $200$ MeV, meanwhile it is constantly set to be zero for $\nu_{\mu}$. Anti-neutrinos to each species are assumed to have the same spectrum as that of neutrinos, although there are some quantitative difference between them in reality\footnote{This is due to the fact that the neutrino-matter reaction rates are, in general, different between neutrinos and the anti-partners: for instance, the effect of weak magnetism \citep{2002PhRvD..65d3001H}. For more quantitative arguments, the difference should be taken into account, which is beyond the scope of this paper, though.}.

In our analytic formula, we have four independent free parameters: $A, B, T,$ and $Q$. We first determine the parameters associated with the thermal component, i.e., $A$ and $T$ by referring some CCSN simulations \citep{2019ApJS..240...38N,2019MNRAS.490.4622N}. For early post-bounce phase ($\lesssim 50$ ms after bounce), the energy-luminosity and the average energy for heavy leptonic neutrinos are $\sim 2 \times 10^{52} {\rm erg}/s$ and $\sim 16$ MeV, respectively. We note that the non-thermal component is subdominant contribution to these energy-integrated quantities. From the average energy, we obtain $T \sim 5$ MeV. We note that our Monte carlo simulations presented in the previous section injected neutrinos as a thermal spectrum with $T=6$ MeV, although the emergent spectrum can be well fit by $T \sim 5$ MeV. The energy loss of the thermal component is attributed to the fact that neutrinos with the energy of $\sim 30$ MeV experience non-isoenergetic scatterings with matter (e.g., nucleon recoils and electron-scatterings) during the propagation in post-shock region (see also Fig.~\ref{fig:spec1}). This indicates that our Monte Carlo simulations reproduce the results of CCSN simulations qualitatively; hence we set $T = 5$ MeV as a representative value for the early post-bounce phase. On the other hand, the typical dynamical timescale at this phase is $\Delta T \sim 10$ ms, which determines the coefficient $A$ so as to reproduce the total energy of $L \times \Delta T = 2 \times 10^{50} {\rm erg}$.

\begin{figure*}
\centering
\includegraphics[width=1.0\textwidth]{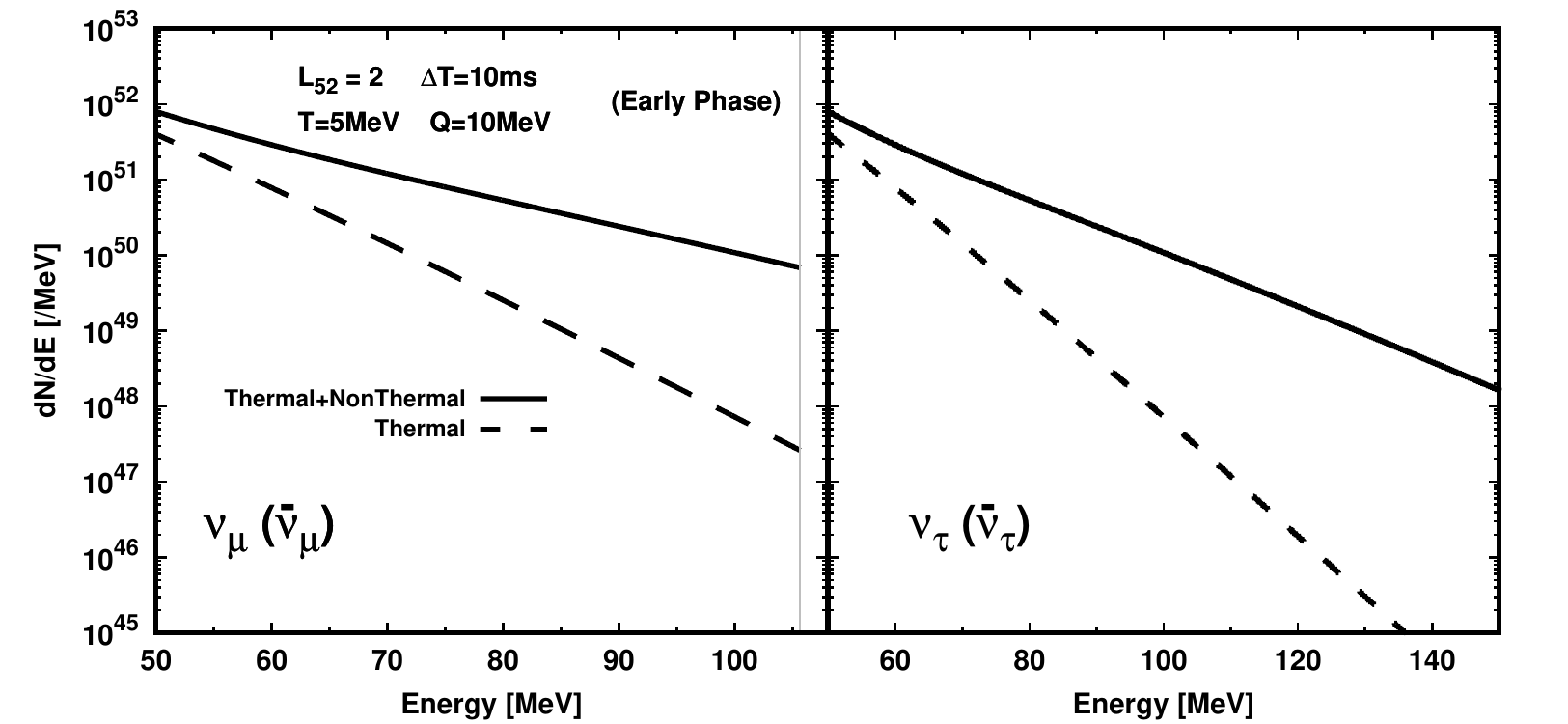}
\caption{Neutrino spectra at the CCSN source modeled by our analytic formula (Eq.~\ref{eq:spectfit_tot}). The left and right panel correspond to $\nu_{\mu}$ and $\nu_{\tau}$, respectively. The parameters are chosen so as to reproduce our Monte Carlo simulations for the early post bounce phase (see text for more details). The solid lines represent the sum of thermal- and non-thermal component of neutrino spectrum. The dashed lines denote those of the thermal component. For $\nu_{\mu}$, the spectrum is cut at the energy of muon rest mass ($106\,{\rm MeV}$) where we draw a thin vertical line in the left panel.}\label{graph_neutrinospectrumSource}
\end{figure*}

We then determine the other two parameters: $B$ and $Q$. In our Monte-Carlo simulations, the exponential decline of non-thermal component can be fit by $Q \sim 10$ MeV. On the other hand, $B$ is determined through the neutrino energy of $\varepsilon$ at which the non-thermal component becomes the same contribution with the thermal one in the spectrum, i.e., 
\begin{eqnarray}
A \hspace{1mm}  \frac{E^2}{1+{\rm exp}( \varepsilon/T  )} = B \hspace{1mm}  E^2  {\rm exp}( -\varepsilon/Q  ) \label{eq:epsiBrela}.
\end{eqnarray}
We find that $\varepsilon = 50$ MeV can reproduce the result of our Monte Carlo simulations. The resultant number spectra of $\nu_{\mu}$ and $\nu_{\tau}$ are displayed in Fig.~\ref{graph_neutrinospectrumSource}.

\begin{figure*}
\centering
\includegraphics[width=1.0\textwidth]{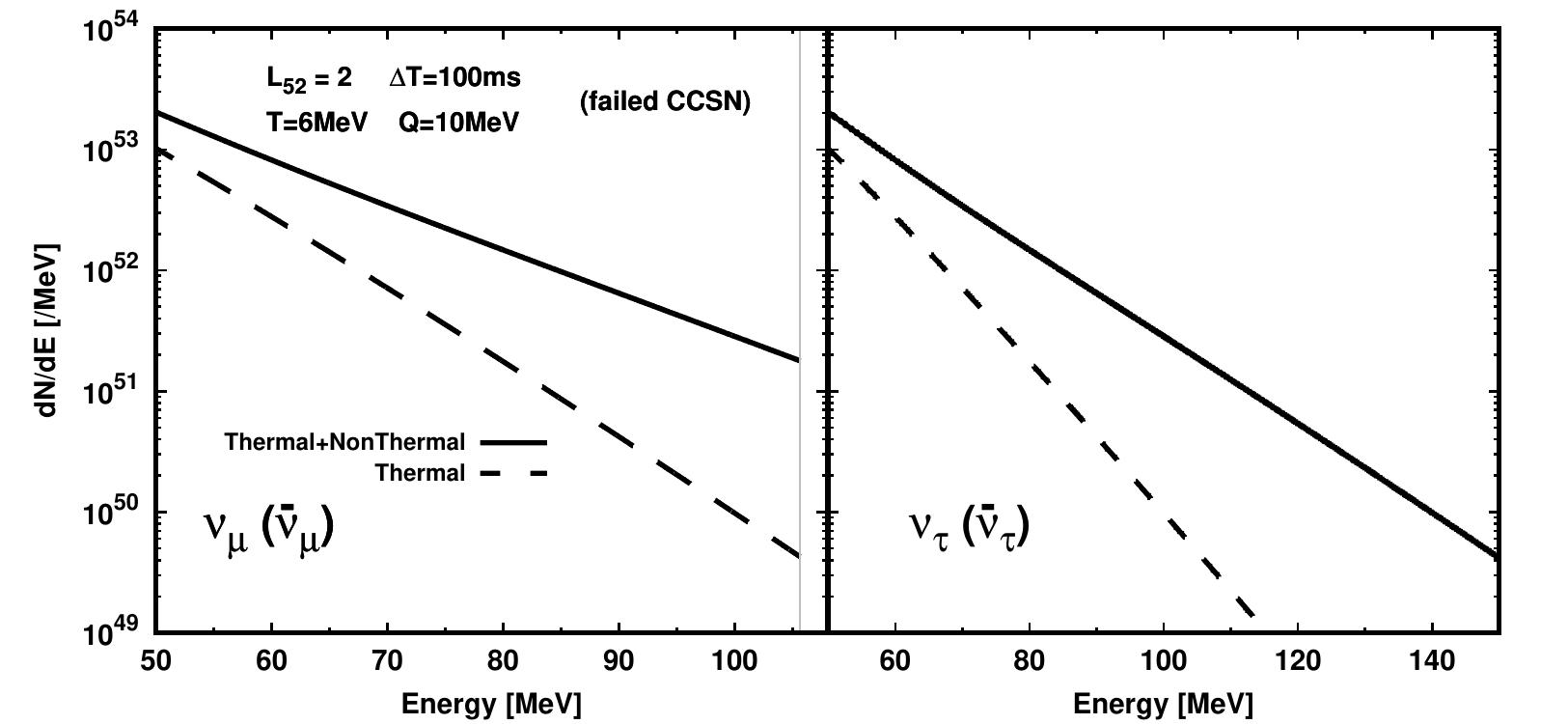}
\caption{Same as Fig.~\ref{graph_neutrinospectrumSource} but for the late post bounce phase in failed CCSN.}\label{graph_neutrinospectrumSource_BH}
\end{figure*}

For the late post bounce phase in failed CCSN, the average energy of heavy leptonic neutrinos tends to be higher than that of the early post-bounce phase \citep[see, e.g.,][]{2020PhRvD.101l3013W}; we set $T = 6$ MeV as a reference. Note that our Monte Carlo simulations also show that the injected thermal spectrum experiences less non-isoenergetic scatterings in the post-shock regions than early post-bounce phase. As a result, the thermal component of the emergent spectrum sustains $T \sim 6$ MeV. We also find that the exponential decline of non-thermal tail depends on the mass accretion rate ($\dot{M}$) but we find $Q > 10$ MeV for $\dot{M} > 1 M_{\sun}$/s; we hence set $Q = 10$ MeV as a conservative choice. Different from the early post bounce phase, on the other hand, the neutrino luminosity varies among models and also depends on time, but the expected value is order of $10^{52} {\rm erg/s}$ \citep[see, e.g.,][]{2006PhRvL..97i1101S,2009A&A...499....1F,2012ApJ...745..197N,2018MNRAS.477L..80K,2020PhRvD.101l3013W}. Just for simplicity, we adopt $L = 2 \times 10^{52} {\rm erg/s}$ as a reference value, which is the same as that of the early post-bounce phase. We note that our choice of the parameters seems to be a conservative, since most of the simulations show that the luminosity is $\gtrsim 2 \times 10^{52} {\rm erg/s}$ and the average energy is also higher than that obtained from $T = 6$ MeV. The dynamical time scale of the system is $\sim 100$ ms; hence we adopt $\Delta T = 100$ ms in this study. Consequently, the total emitted neutrino energy becomes $2 \times 10^{51} {\rm erg}$, which determines $A$ in Eq.~\ref{eq:spectfit_th}. We also find that the Eq.~\ref{eq:epsiBrela} with $\varepsilon = 50$ MeV well reproduces our Monte Carlo simulations, which determines $B$ as the same process in the early post bounce phase. Fig.~\ref{graph_neutrinospectrumSource_BH} portrays the resultant spectrum of heavy leptonic neutrinos for the late post bounce phase in failed CCSN.

We must mention several caveats regarding our choice of the parameters. Although the choice was made based on the emergent spectra obtained by our Monte Carlo simulations, there remain several uncertainties, indicating that the sensitivity of the detectability to the parameters needs to be investigated. As we shall show below, however, that there also remain large uncertainties in neutrino cross sections with detector materials, which prevents the quantitative arguments; hence, our discussions are restricted to a qualitative level. We postpone the detailed study of parameter dependence in future until we remove or at least reduce the major uncertainties for the estimation.

\subsubsection{Neutrino oscillation}\label{subsubsec:neutrinoosci}
As we have described in Sec.~\ref{sec:theoryshockac}, the neutrino shock acceleration breaks the degeneracy of $\nu_{\mu}$ and $\nu_{\tau}$ in the energy of $E > M_{u}$, implying that the treatment of three flavor of neutrinos is indispensable. Three different flavors of neutrinos change into each other during flight due to neutrino oscillation, which should be taken into account to consider the event count in terrestrial detectors. In this paper, we adopt a simple oscillation model but frequently used in the literature: adiabatic Mikheyev-Smirnov-Wolfenstein (MSW) model for normal and inverted mass hierarchies. Below, we describe the essence of the model.

The CCSN core is the place where the matter potential of the neutrino oscillation Hamiltonian dominates the vacuum one. The matter potential is not identical among different flavors; for instance, charged-current interactions in $\nu_e$ make the matter potential higher than that for other heavy leptonic neutrinos. We also note that the radiative corrections in matter reactions depend on the mass of leptons \citep{1987PhRvD..35..896B}, indicating that $\nu_{\mu}$ and $\nu_{\tau}$ also feel the different matter potential. Although the radiative correction is much smaller than the charged-current interactions, the difference plays an important role to distinguish $\nu_{\mu}$ and $\nu_{\tau}$, and in particular, the effect overwhelms the vacuum potential if the matter density ($\rho$) becomes higher than $\sim 10^{7} - 10^{8} {\rm g/cm}^3$ \citep{1987PhRvD..35..896B,2000PhRvD..62c3007D}. We find that the neutrino shock acceleration occurs at the place where the matter density is comparable or higher than the threshold; hence, we assume that the three flavors of neutrinos are pinned at each three different mass eigenstate in this study.

To see the relation between the flavor- and effective mass eigenstate of neutrinos in matter, we need to compute the eigenvalues of the oscillation Hamiltonian. For neutrinos, the Hamiltonian in the flavor basis can be written as,
\begin{eqnarray}
\mathbfss{H} = \mathbfss{H}_{\rm v} + \mathbfss{H}_{\rm m}, \label{eq:HamiltNeut}
\end{eqnarray}
where
\begin{eqnarray}
\mathbfss{H}_{\rm v} = \frac{1}{2E} \mathbfss{U}
\begin{pmatrix}
m_1^2 & 0 & 0 \\
0 & m_2^2 & 0 \\
0 & 0 & m_3^2
\end{pmatrix} \mathbfss{U}^{\dagger},
  \label{eq:HamiltNeut_vac}
\end{eqnarray}
and
\begin{eqnarray}
\mathbfss{H}_{\rm m} =
\begin{pmatrix}
V_{e \mu} & 0 & 0 \\
0 & 0 & 0 \\
0 & 0 & V_{\tau \mu}
\end{pmatrix}.
  \label{eq:HamiltNeut_mat}
\end{eqnarray}
In the expressions, $m_i$ ($i=1, 2, 3$) denotes the three independent mass of neutrinos. $\mathbfss{U}$ represents the  Pontecorvo-Maki-Nakagawa-Sakata (PMNS) matrix\footnote{We ignore the two Majorana phases in the PMNS matrix, since they do not affect neutrino oscillations \citep{1980PhLB...94..495B,1987NuPhB.282..589L}.},
\begin{eqnarray}
\mathbfss{U} = \mathbfss{U}_{23} \mathbfss{U}_{13} \mathbfss{U}_{12}
  \label{eq:PMNStot}
\end{eqnarray}
where
\begin{eqnarray}
&&\mathbfss{U}_{23} =
\begin{pmatrix}
1 & 0 & 0 \\
0 & c_{23} & s_{23} \\
0 & -s_{23} & c_{23}
\end{pmatrix},  \label{eq:PMNStot23} \nonumber \\
&&\mathbfss{U}_{13} =
\begin{pmatrix}
c_{13} & 0 & s_{13} \hspace{0.5mm} e^{-i \delta_{cp}} \\
0 & 1 & 0 \\
-s_{13} \hspace{0.5mm} e^{i \delta_{cp}} & 0 & c_{13}
\end{pmatrix},  \label{eq:PMNSto13} \nonumber \\
&&\mathbfss{U}_{12} =
\begin{pmatrix}
c_{12} & s_{12} & 0 \\
-s_{12} & c_{12} & 0 \\
0 & 0 & 1
\end{pmatrix}.  \label{eq:PMNSto12}
\end{eqnarray}
$c_{ij}$ and $s_{ij}$ are $\cos \theta_{ij}$ and $\sin \theta_{ij}$, respectively ($\theta_{ij}$ denotes the neutrino mixing angles), and $\delta_{cp}$ denotes the CP violation phase. $V_{e \mu}$ and $V_{\tau \mu}$ denote the matter potential with respect to $\nu_{e}$ and $\nu_{\tau}$, respectively\footnote{In the expression, we subtracted the contribution of matter potential with respect to $\nu_{\mu}$.}, which can be written as
\begin{eqnarray}
V_{e \mu} \sim \sqrt{2} G_{F} n_e, \label{eq:Vtaumu}
\end{eqnarray}
where $n_e$ denotes the number density of electron, and
\begin{eqnarray}
V_{\tau \mu} \sim 10^{-4} V_{e \mu}, \label{eq:Vtaumu}
\end{eqnarray}
for the case with $n_e \sim n_p \sim n_n$\footnote{It is a reasonable condition, since the electron fraction around the shock radius is $\sim 0.5$.}, where $n_p$ and $n_n$ denotes the number density of free proton and neutron, respectively \citep[see also][for more complete descriptions of $V_{\tau \mu}$]{1987PhRvD..35..896B,2000PhRvD..62c3007D}.

There are three independent eigenvalues of $\mathbfss{H}$, which can be written as
\begin{eqnarray}
\lambda_k = - \frac{b}{3} + \frac{2}{\sqrt{3}} \sqrt{-p} \cos \left(
\frac{1}{3} \arccos( \frac{3 \sqrt{3} q  }{2p \sqrt{-p}} ) + \frac{2 \pi}{3} k
\right) \hspace{2mm}, \label{eq:EigenofH}
\end{eqnarray}
where $k$ runs from 1 to 3. In the expression,
\begin{eqnarray}
b = - \left(  m_1^{*2} + m_2^{*2} + m_3^{*2} + V_{\tau \mu} + V_{e \mu} \right),
\label{eq:coefbdef}
\end{eqnarray}
where
\begin{eqnarray}
m_i^{*2} \equiv \frac{m_i^2}{2E} \label{eq:mstardef},
\end{eqnarray}
and
\begin{eqnarray}
&&p = - \frac{b^2}{3} + c, \nonumber \\
&&q = \frac{2}{27} b^3 - \frac{1}{3} bc + d, \label{eq:pqcoef}
\end{eqnarray}
where
\begin{eqnarray}
c &=& m_1^{*2} m_2^{*2} + m_1^{*2} m_3^{*2} + m_2^{*2} m_3^{*2} \nonumber \\
 &+& \left(  ( \mathbfss{H}_{\rm v} )_{ee} + ( \mathbfss{H}_{\rm v} )_{\mu \mu} \right) V_{\tau \mu} \nonumber \\
  &+& \left( ( \mathbfss{H}_{\rm v} )_{\mu \mu} + ( \mathbfss{H}_{\rm v} )_{\tau \tau}
\right) V_{e \mu}, \label{eq:defccoef} \\
d &=& - \{\{ 
m_1^{*2} m_2^{*2} m_3^{*2} + ( \mathbfss{H}_{\rm v} )_{e e} ( \mathbfss{H}_{\rm v} )_{\mu \mu} V_{\tau \mu} \nonumber \\
&+&  \left[ ( \mathbfss{H}_{\rm v} )_{\mu \mu} ( \mathbfss{H}_{\rm v} )_{\tau \tau}
                    + ( \mathbfss{H}_{\rm v} )_{\mu \mu} V_{\tau \mu}
                    - ( \mathbfss{H}_{\rm v} )_{\tau \mu} ( \mathbfss{H}_{\rm v} )_{\mu \tau} 
  \right] V_{e \mu} \nonumber \\
&-& ( \mathbfss{H}_{\rm v} )_{e \mu} ( \mathbfss{H}_{\rm v} )_{\mu e} V_{\tau \mu}   
\}\}. \label{eq:defdcoef} 
\end{eqnarray}
For the case with normal mass hierarchy ($m_3 > m_2 > m_1$), the three eigenvalues in Eq.~\ref{eq:EigenofH} correspond to the effective mass eigenstates as,
\begin{eqnarray}
&&\lambda_0: \nu_3  , \nonumber \\
&&\lambda_1: \nu_1  , \nonumber \\
&&\lambda_2: \nu_2  , \label{eq:eigencorespmassNM}
\end{eqnarray}
respectively.
By taking the high density limit, we obtain
\begin{eqnarray}  
&&\lambda_0 \sim V_{e \mu} , \nonumber \\
&&\lambda_2 \sim V_{\tau \mu} , \label{eq:eigenhighdenslimNM}
\end{eqnarray}
and then we obtain
\begin{eqnarray}
&&\nu_e \sim \nu_3  , \nonumber \\
&&\nu_{\mu} \sim \nu_1  , \nonumber \\
&&\nu_{\tau} \sim \nu_2  , \label{eq:FlavorMassCoNM}
\end{eqnarray}
For the case with inverted mass hierarucy, each $\lambda$ corresponds to the effective mass eigenstate as
\begin{eqnarray}
&&\lambda_0: \nu_2 , \nonumber \\
&&\lambda_1: \nu_3 , \nonumber \\
&&\lambda_2: \nu_1 , \label{eq:eigencorespmassInV}
\end{eqnarray}
respectively. The two of them coincide with the matter potentials in high density limit as
\begin{eqnarray}  
&&\lambda_2 \sim V_{e \mu} , \nonumber \\
&&\lambda_1 \sim V_{\tau \mu}, \label{eq:eigenhighdenslimInV}
\end{eqnarray}
hence we obtain
\begin{eqnarray}
&&\nu_e \sim \nu_2  , \nonumber \\
&&\nu_{\mu} \sim \nu_3  , \nonumber \\
&&\nu_{\tau} \sim \nu_1. \label{eq:FlavorMassCoInV}
\end{eqnarray}

For anti-neutrinos, we replace the Hamiltonian as
\begin{eqnarray}
&&\mathbfss{H}_{\rm v} \rightarrow \mathbfss{H}^{*}_{\rm v}, \nonumber \\
&&\mathbfss{H}_{\rm m} \rightarrow - \mathbfss{H}_{\rm m},
\end{eqnarray}
in Eq.~\ref{eq:HamiltNeut}. By taking the same process as neutrinos, we obtain
\begin{eqnarray}
&&\bar{\nu}_e \sim \bar{\nu}_1  , \nonumber \\
&&\bar{\nu}_{\mu} \sim \bar{\nu}_3  , \nonumber \\
&&\bar{\nu}_{\tau} \sim \bar{\nu}_2  , \label{eq:FlavorMassCoNM_anti}
\end{eqnarray}
for normal mass hierarchy, and
\begin{eqnarray}
&&\bar{\nu}_e \sim \bar{\nu}_3  , \nonumber \\
&&\bar{\nu}_{\mu} \sim \bar{\nu}_2  , \nonumber \\
&&\bar{\nu}_{\tau} \sim \bar{\nu}_1  , \label{eq:FlavorMassCoInV_anti}
\end{eqnarray}
for inverted mass hierarchy. Following the adiabatic MSW model, we assume that neutrinos and anti-neutrinos arrive at terrestrial detectors with the same mass eigenstate when they were in CCSN. Consequently, we obtain the neutrino number flux of each flavor state at detectors as,
\begin{eqnarray}
F_{\alpha} = \sum_i^{3} | U_{\alpha i} |^2  F_{i}, \nonumber \\
\bar{F}_{\alpha} = \sum_i^{3} | U^{*}_{\alpha i} |^2  \bar{F}_{i}, \label{eq:conversion}
\end{eqnarray}
where $F_{\alpha}$ and $F_{i}$ denote neutrino spectrum in flavor- and mass- eigenstate, respectively. Regarding neutrino mixing parameters embedded in PMNS matrix, we adopt the current best fit values of $\theta_{ij}$ and $\delta_{cp}$ provided by NuFIT 5.0 with SK atmospheric data \citep{2020arXiv200714792E}.

It should be mentioned that the neutrino oscillation model adopted in this study is the simplest one and need to be improved for more detailed estimation. For instances, the Earth matter effect should be taken into account for some detectors in reality \citep{2001NuPhB.616..307L}. Neutrino-neutrino self-interaction in the oscillation Hamiltonian also needs to be incorporated; indeed, \citet{2020PhRvR...2a2046M} recently revealed that the fast pair-wise conversion commonly occurs in the pre-shock region at $\gtrsim 100$ ms after bounce, implying that it affects the event rate in the late post bounce phase for failed CCSN. The recent CCSN studies also suggest that the multi-dimensional fluid instabilities in post-shock flows assist the occurrence of fast pairwise conversions \citep{2019PhRvD.100d3004A,2019ApJ...886..139N,2020PhRvD.101b3018D}. Unfortunately, however, the final outcome of the flavor conversion due to neutrino self interactions is still elusive, although the community has recently made extensive efforts to address the issue \citep[see, e.g.,][]{2019ApJ...883...80S,2019PhRvD..99l3014R,2020PhRvD.101f3027S,2020PhRvD.101d3009J,2020PhRvD.101f3001G,2020JCAP...05..027A,2020arXiv200500459B,2020arXiv200909024J,2020arXiv200903337B}. From a different point of view, on the other hand, the detection of high energy neutrinos in real observations will be a valuable data to place a constraint on such a complex neutrino oscillation model, which will be also discussed in Sec.~\ref{subsec:muondetec}.

\subsubsection{Detector configurations}\label{subsubsec:detectorconfig}
We employ four representative terrestrial neutrino detectors: SK, HK, DUNE, and JUNO, to estimate the event count. Just for simplicity, we do not take into account any smearing effects by detector responses and noises in this study, although they should be taken into account in real observations. Below, we provide some essential information on each detector configuration.

We assume that SK and HK have the identical detector configuration except for the fiducial volume. The detector scale of the former and latter is assumed to be 32.5 ktons \citep{2016APh....81...39A} and 220 ktons \citep{2018arXiv180504163H}, respectively. It should be noted that we only focus on charged-current reaction channels in this study, although those of neutral current would also provide useful information to study the neutrino oscillation model (see below for more details). One of the primary charged current reactions in SK and HK is the inverse beta decay on proton for $\bar{\nu}_e$ (IBD-p-nueb)
\begin{eqnarray}
&&\bar{\nu}_{e}  + p \rightarrow {e}^{+} + n.
\label{ibdchart_nueb}
\end{eqnarray}
For the energy of $\gtrsim 106\,{\rm MeV}$, the same reaction but for $\bar{\nu}_{\mu}$ (IBD-p-numub) also appears,
\begin{eqnarray}
&&\bar{\nu}_{\mu}  + p \rightarrow {\mu}^{+} + n,
\label{ibdchart_numub}
\end{eqnarray}
which is also taken into account in this study. For high energy neutrinos ($\gtrsim 50$ MeV), the charged current reactions with Oxygen:
\begin{eqnarray}
&&\nu_e  + {^{16}{\rm O}}  \rightarrow e^{-} + {^{16}{\rm F}}\, , \label{eq:Onue} \\
&&\bar{\nu}_e  + {^{16}{\rm O}}  \rightarrow e^{+} + {^{16}{\rm N}}\, \label{eq:Onueb},
\end{eqnarray}
become important, and their muon-channels:
\begin{eqnarray}
&&\nu_{\mu}  + {^{16}{\rm O}}  \rightarrow {\mu}^{-} + {^{16}{\rm F}}\, , \label{eq:Onumu} \\
&&\bar{\nu}_{\mu}  + {^{16}{\rm O}}  \rightarrow {\mu}^{+} + {^{16}{\rm N}}\, \label{eq:Onumub},
\end{eqnarray}
also appear in $\gtrsim 106\,{\rm MeV}$. In total, six independent channels are adopted for SK and HK.

We employ a publicly available cross section data in SNOwGLoBES\footnote{The software is available from \url{https://webhome.phy.duke.edu/~schol/snowglobes/}. Note that we do not use the analysis pipeline, since it is only available to estimate the event count up to $100$ MeV.}. However, the data is insufficient; for instances, most of them are available up to $100 {\rm MeV}$ and those of muon channels are not provided\footnote{We note that the extended data for the cross sections will be available in SNOwGLoBES (Kate Scholberg and Joel Dai, private communications).}. We, hence, complement the cross section data by collecting other references with some extensions.

For the IBD-p-nueb cross section implemented in SNOwGLoBES, it is available up to $200$ MeV, which is computed by \citet{2003PhLB..564...42S}. For the IBD-p-numub cross section, we employ a fitting formula in \citet{1980PhRvD..21..562F} with a modification, which is
\begin{eqnarray}
\sigma = 0.32 \times 10^{-38} \left( \frac{E - E_{\rm thre}}{1 {\rm GeV}} \right) \hspace{2mm} [{\rm cm}^2], \label{eq:fitformulaofnumubp}
\end{eqnarray}
where $E_{\rm thre} \equiv M_{n} - M_{p} + M_{\mu}$ ($M_{n}, M_{p}, M_{\mu}$ denote the rest mass energy of neutron, proton and muon, respectively), and we assume $\sigma = 0$ in the energy of $E \leq E_{\rm thre}$. For reactions with $\nu_e-{^{16}{\rm O}}$ and $\bar{\nu}_e-{^{16}{\rm O}}$, the cross sections are available up to $100$ MeV in SNOwGLoBE; hence we fit them quadratically in the energy range between $50$ MeV and $100$ MeV, and then extrapolates the cross section up to $200$ MeV. The cross section data for $\nu_{\mu}-{^{16}{\rm O}}$ and $\bar{\nu}_{\mu}-{^{16}{\rm O}}$ are taken from \citet{2012RvMP...84.1307F}.

For DUNE, we assume that the detector scale is $40$ ktons and the four charged current reactions are taken into account in this study,
\begin{eqnarray}
&&\nu_e  + {^{40}{\rm Ar}} \rightarrow e^{-} + {^{40}{\rm K}^{*}}\, , \label{eq:CCArnue} \\
&&\bar{\nu}_e  + {^{40}{\rm Ar}} \rightarrow e^{+} + {^{40}{\rm Cl}^{*}}\, , \label{eq:CCArnueb} \\
&&\nu_{\mu}  + {^{40}{\rm Ar}} \rightarrow {\mu}^{-} + {^{40}{\rm K}^{*}}\, , \label{eq:CCArnumu} \\
&&\bar{\nu}_{\mu}  + {^{40}{\rm Ar}} \rightarrow {\mu}^{+} + {^{40}{\rm Cl}^{*}}\, . \label{eq:CCArnumub} 
\end{eqnarray}
The cross section data of $\nu_e-{^{40}{\rm Ar}}$ and $\bar{\nu}_e-{^{40}{\rm Ar}}$ are available up to $200 {\rm MeV}$ in SNOwGLoBES, which are computed in \citet{2004JCAP...08..001G,2003JPhG...29.2569K}. For those of the $\nu_{\mu}$ and $\bar{\nu}_{\mu}$ reactions, we use the result in \citet{2015IJMPE..2450079A}, which provides the ratio of the cross sections between $\nu_e$ ($\bar{\nu}_e$) and $\nu_{\mu}$ ($\bar{\nu}_{\mu}$). We employ the result of the Local Fermi Gas Model with taking into account the effect of nuclear excitation under a random phase approximation, which can be approximately given as
\begin{eqnarray}
\left( \frac{\sigma_{\mu}}{\sigma_e} \right)_{\rm Ar} = 0.6 \hspace{0.5mm} \frac{( E - M_{\mu})}{( 200 {\rm MeV} - M_{\mu})} , \label{eq:RatioLambda_forArgon}
\end{eqnarray}
up to $200$ MeV and it is assumed to be zero for $E < M_{\mu}$. The ratio is nearly the same in anti-neutrinos; hence Eq.~\ref{eq:RatioLambda_forArgon} is adopted for them. Based on these assumptions, we can evaluate the cross sections of the reactions with Eqs.~\ref{eq:CCArnumu}~and~\ref{eq:CCArnumub} from those of $\nu_e$ and $\bar{\nu}_e$ (Eqs.~\ref{eq:CCArnue}~and~\ref{eq:CCArnueb}, respectively).

For JUNO, we assume that the fiducial volume is $20$ ktons. The detector employs Linear Alkyl Benzene for the liquid scintillation system \citep{2017JInst..1211004Y}; hence the mass fraction of Carbon and Hydrogen is $88 \%$ and $12 \%$, respectively \citep[see also][]{2016JPhG...43c0401A}. We employ IBD-p-nueb and IBD-p-numub channels (similar as SK and HK). Other four charged current reaction channels with Carbon are also taken into account in this study,
\begin{eqnarray}
&&\nu_e  + {^{12}{\rm C}}  \rightarrow e^{-} + {^{12}{\rm N}}\, , \label{eq:Cnue} \\
&&\bar{\nu}_e  + {^{12}{\rm C}}  \rightarrow e^{+} + {^{12}{\rm B}}\, \label{eq:Cnueb}, \\
&&\nu_{\mu}  + {^{12}{\rm C}}  \rightarrow {\mu}^{-} + {^{12}{\rm N}}\, , \label{eq:Cnumu} \\
&&\bar{\nu}_{\mu}  + {^{12}{\rm C}}  \rightarrow {\mu}^{+} + {^{12}{\rm B}}\, \label{eq:Cnumub}.
\end{eqnarray}
For the cross section of $\nu_e-{^{12}{\rm C}}$ and $\bar{\nu}_e-{^{12}{\rm C}}$, we adopt the data computed by \citet{1999NuPhA.652...91K}. For those of $\nu_{\mu}$, we take the same prescription as Argon (see Eqs.~\ref{eq:RatioLambda_forArgon}), although the low energy threshold is slightly higher for Carbon than that of Argon \citep[see, e.g.,][]{2017JPhG...44l5108A}, which is $\sim 120 {\rm MeV}$. For this reason, we employ the following formula,
\begin{eqnarray}
\left( \frac{\sigma_{\mu}}{\sigma_e} \right)_{\rm C} = 0.4 \hspace{0.5mm} \frac{(E - 120 {\rm MeV} )}{80 {\rm MeV}}, \label{eq:RatioLambda_forC}
\end{eqnarray}
to compute the cross section of charged current reactions of $\nu_{\mu}$ with Carbon.

As described above, some of the neutrino cross sections with detector materials are estimated with artificial prescriptions, implying that they are inaccurate. As is well known, however, the cross sections are less constrained by both experiments and theories, indicating that no definitive data are available at the moment. The uncertainties may be the largest obstacles to develop more quantitative arguments for the detectability of the high energy neutrinos. Although addressing the issue is much beyond the scope of this paper, we need to keep in mind the uncertainty to interpret our results presented in the following sections.

\subsection{Detectability of high energy neutrinos in early post bounce phase}\label{subsec:deteearlyresults}

\begin{figure*}
\centering
\includegraphics[width=1.0\textwidth]{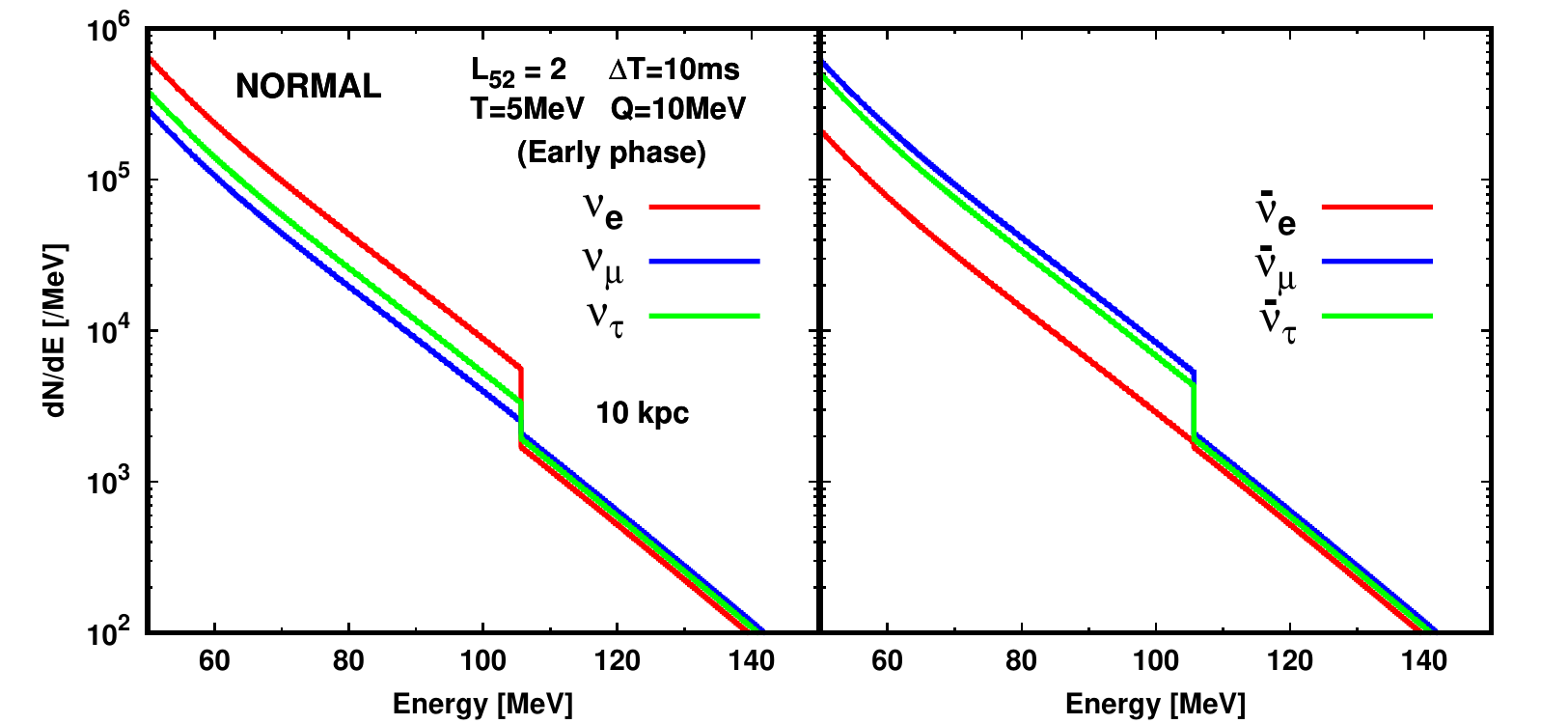}
\caption{Neutrino spectra measured at the Earth in early post bounce phase. The CCSN source distance is assumed to be $10 {\rm kpc}$. The left and right panels show the spectra for neutrinos and anti-partners, respectively. The color represents the flavor of neutrinos. The flavor conversion is assumed as an adiabatic MSW model with normal mass hierarchy (see Sec.~\ref{subsubsec:neutrinoosci}). For the detail of our analytic formula, see text and equations in Sec.~\ref{subsubsec:anaexpress}.}\label{graph_neutrinospectrum_normal}
\end{figure*}

\begin{figure*}
\centering
\includegraphics[width=1.0\textwidth]{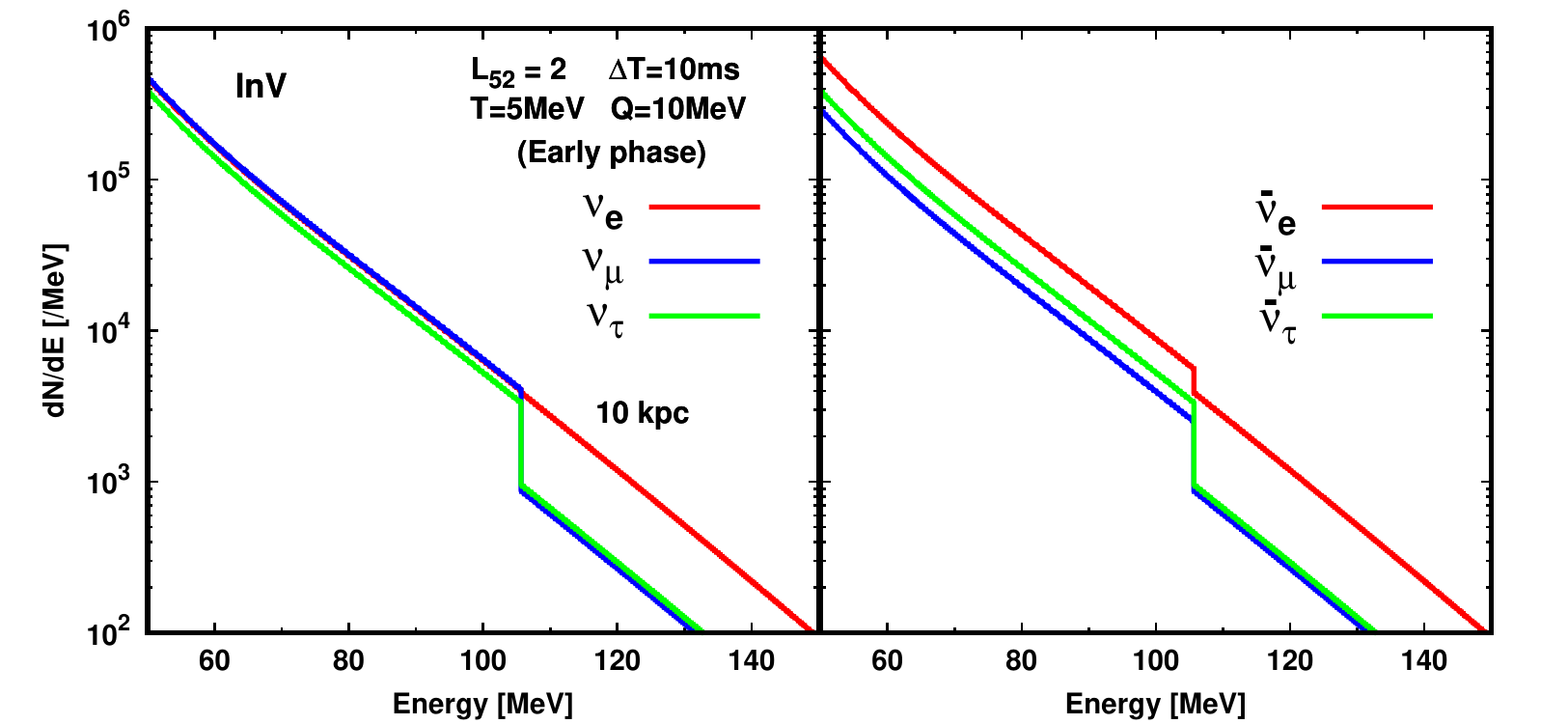}
\caption{Same as Fig.~\ref{graph_neutrinospectrum_normal} but for the case with inverted mass hierarchy.}\label{graph_neutrinospectrum_inv}
\end{figure*}

We present the result of our estimation with focusing on the early post bounce phase. Unless otherwise states, the distance to the CCSN source is assumed to be $10$ kpc. We first show the neutrino spectra measured at the Earth in Figs.~\ref{graph_neutrinospectrum_normal}~and~\ref{graph_neutrinospectrum_inv} for normal mass hierarchy and inverted mass hierarchy, respectively. It should be emphasized that the appearance of $\nu_e$ and $\bar{\nu}_e$ is a consequence of neutrino flavor conversion (since they are assumed to be zero at the CCSN source). The discrete change in the spectra at $\sim 100$ MeV reflects the disappearance of $\nu_{\mu}$ and $\bar{\nu}_{\mu}$ in the energy of $E>M_u$ at the source (see Sec.~\ref{sec:theoryshockac}). This also implies that the appearance of $\nu_{\mu}$ and $\bar{\nu}_{\mu}$ in the energy of $E>M_u$ at the Earth is another consequence of neutrino flavor conversion; they are originated from $\nu_{\tau}$ and $\bar{\nu}_{\tau}$ accelerated at the shock wave. It opens a new possibility of muon production at detectors; the detail will be discussed in Sec.~\ref{subsec:muondetec}.

\begin{figure*}
\centering
\includegraphics[width=1.0\textwidth]{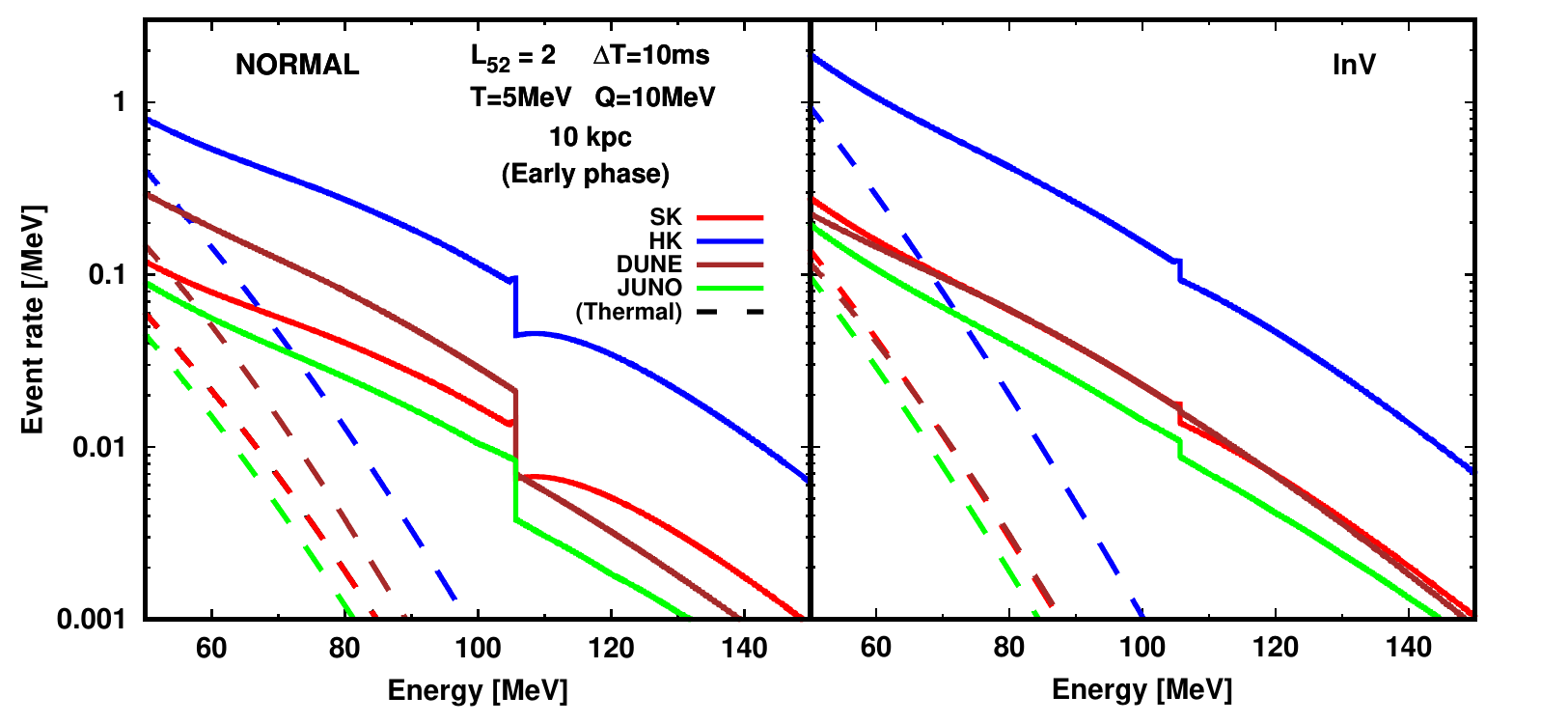}
\caption{Energy spectra of event counts for neutrinos in the early post bounce phase. The color denotes the detector: SK (red), HK (blue), DUNE (brown), and JUNO (light green). The event counts are integrated charged current reaction channels of each detector as described in Sec.~\ref{subsubsec:detectorconfig}. Left and right panels show the results for the normal mass hierarchy and inverted one, respectively. For comparison, we also display the energy spectra of events counts for thermal neutrinos as dashed lines.}\label{graph_Detectionnumspectrum}
\end{figure*}

\begin{figure*}
\centering
\includegraphics[width=1.0\textwidth]{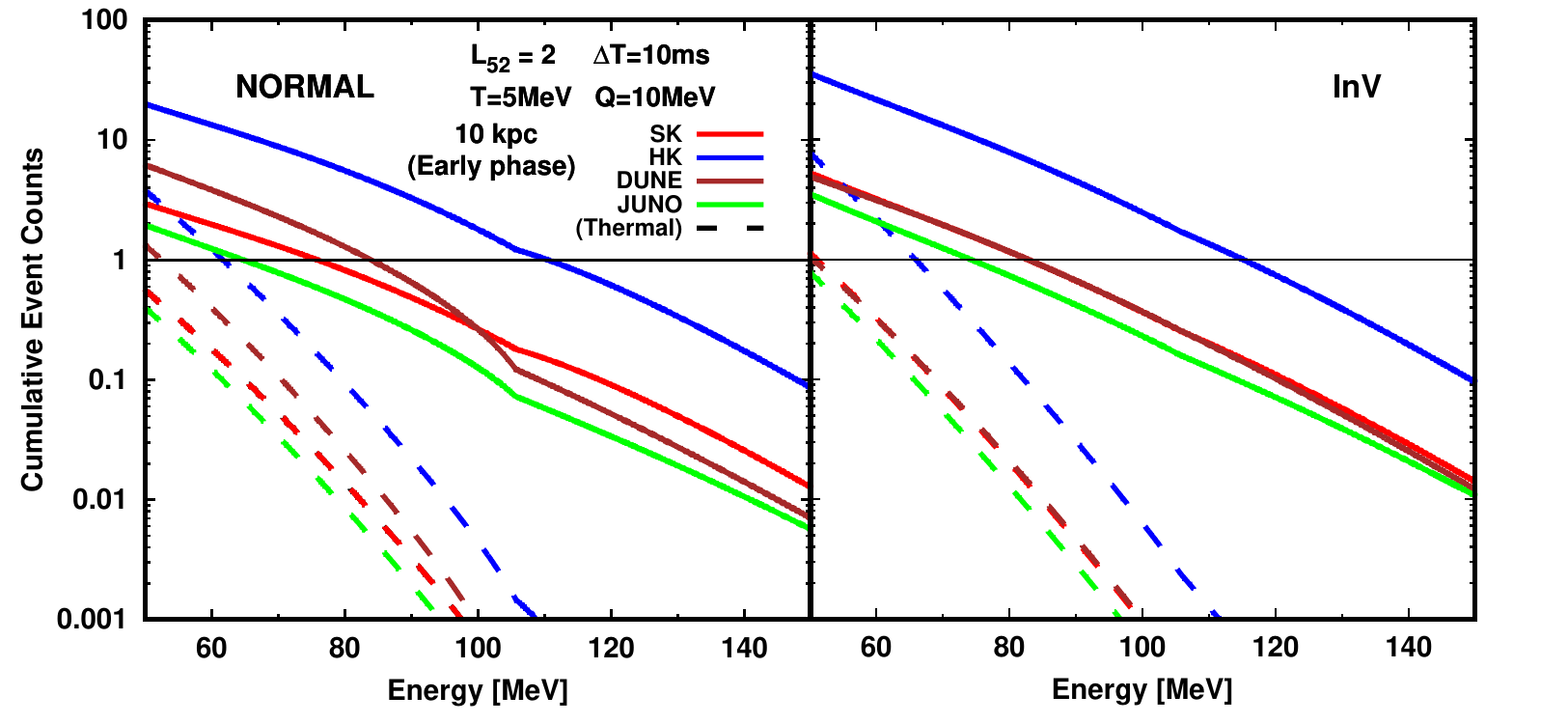}
\caption{Same as Fig.~\ref{graph_Detectionnumspectrum} but for cumulative event counts. The integration starts from $200 {\rm MeV}$ and then proceeds toward the negative energy. The neutrino energy where each line crosses with a thin black line corresponds to the expected maximum energy of neutrinos by each detector. For instance, HK is capable of detecting neutrinos up to $\sim 110$ MeV.}\label{graph_CumulativeFromHigh}
\end{figure*}

\begin{figure*}
\centering
\includegraphics[width=1.0\textwidth]{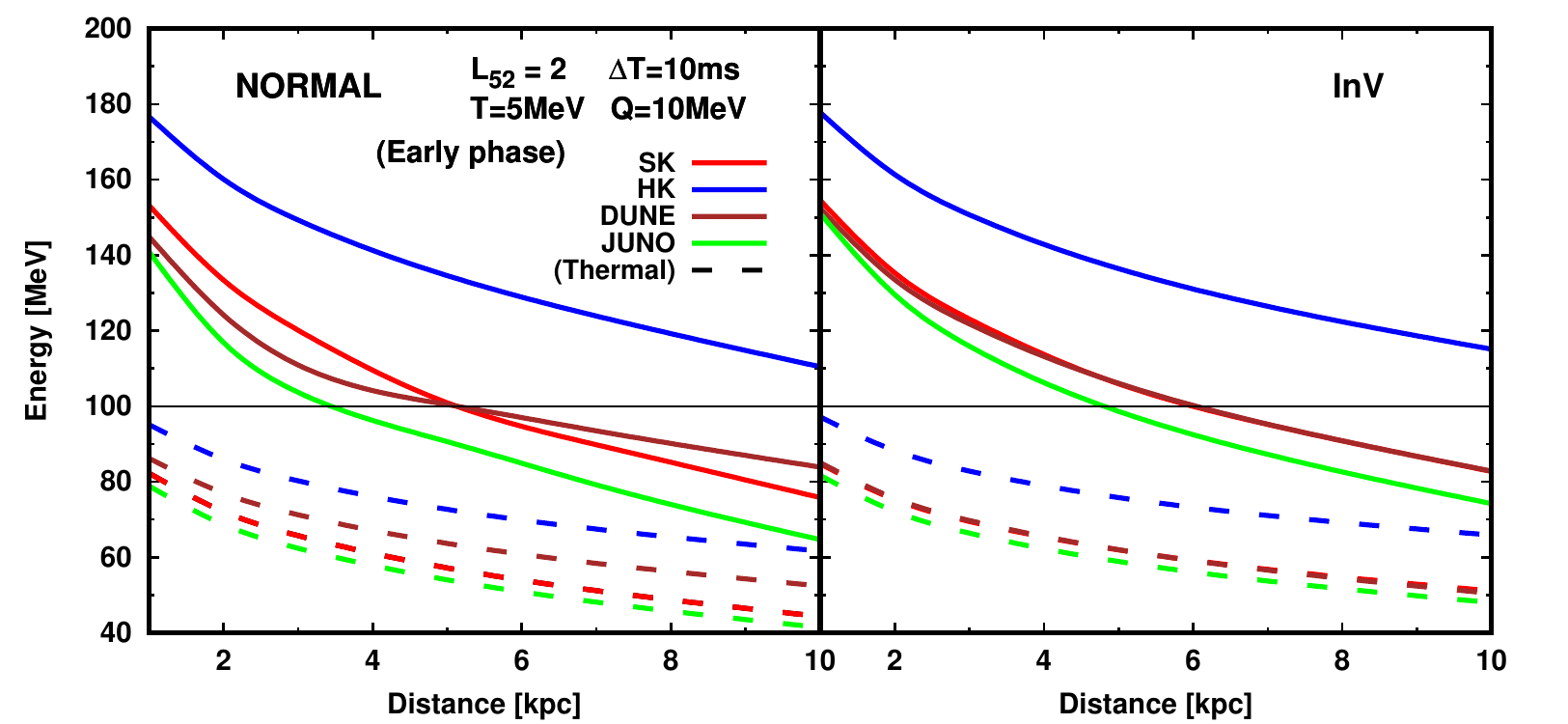}
\caption{The expected maximum energy of detected neutrinos (emitted in the early post bounce phase) via charged current reaction channels as a function of the distance to CCSN. The color and line type denote the same as those used in Fig.~\ref{graph_Detectionnumspectrum}.}\label{graph_Detectionthreshold}
\end{figure*}

The energy spectra of event counts on each detector are displayed in Fig.~\ref{graph_Detectionnumspectrum}. In the spectrum, we sum up the event count for all reaction channels (charged current reactions described in Sec.~\ref{subsubsec:detectorconfig}). We confirm that the event counts at $E \gtrsim 80$ MeV are orders of magnitude higher than those for thermal neutrinos regardless of detectors, and HK provides the largest event counts regardless of neutrino mass hierarchy. We also find that the event count up to $E=M_u$ is higher for DUNE than SK in normal mass hierarchy, meanwhile the difference is less remarkable in inverted one. It is attributed to the fact that DUNE is the most sensitive to $\nu_e$ which is abundant at the Earth in normal mass hierarchy (see the left panel in Fig.~\ref{graph_neutrinospectrum_normal}). It also reflects the fact that the survival probability of $\nu_e$ is nearly zero and substantial heavy leptonic neutrinos at the CCSN source are converted to $\nu_e$, instead. On the other hand, the opposite trend emerges in $E > M_{\mu}$ for normal mass hierarchy; the event count becomes higher in SK than that of DUNE (see the left panel of Fig.~\ref{graph_Detectionnumspectrum}). This is mainly due to the disappearance of $\nu_{\mu}$ at the CCSN source, which reduces the $\nu_e$ flux by a factor of $\sim 3$ at the Earth (see the red line in the left panel of Fig.~\ref{graph_neutrinospectrum_normal}). Although the event count in $\nu_e$ reaction with Oxygen (Eq.~\ref{eq:Onue}) at SK also has an impact of the reduction of $\nu_e$ flux at the Earth, the $\bar{\nu}_e$ at the Earth is less sensitive to the reduction of $\bar{\nu}_{\mu}$ at the CCSN source in normal mass hierarchy (see the red line in the right panel of Fig.~\ref{graph_neutrinospectrum_normal}); as a consequence the reduction in the total event count becomes weaker in SK than that in DUNE.

We display the energy spectrum of cumulative event counts in Fig.~\ref{graph_CumulativeFromHigh}. In the plot, the event count is integrated from $200$ MeV towards the low energy; hence, the spectrum results in monotonically decreasing with neutrino energy. The cumulative event counts defined above are useful to assess the detectability of high energy neutrinos; for instance, it enable us to determine the expected maximum energy of detected neutrinos on each detector. It corresponds to the energy where the cumulative event counts reaches unity, i.e., it is $\sim 110$ MeV, $\sim 76$ MeV, $\sim 84$ MeV, and $\sim 65$ MeV for HK, SK, DUNE, and JUNO, respectively, in normal mass hierarchy; in inverted mass hierarchy, it is $\sim 115$ MeV, $\sim 83$ MeV, $\sim 84$ MeV, and $\sim 74$ MeV in the same order of detectors. It should be stressed that the expected maximum energies are remarkably higher than those of thermal components (see dashed lines in the same figure). We also note that the threshold energy for thermal component is less sensitive to the source distance than that for non-thermal one, which can be seen in Fig.~\ref{graph_Detectionthreshold}. Regardless of the neutrino mass hierarchy and detectors, $>100$ MeV thermal neutrinos are not detectable (see dashed lines in the figure) unless the distance to the CCSN source is very nearby $\lesssim 1$ kpc. The insensitiveness to the source distance reflects an important fact that the exponential decline of the thermal spectrum of neutrinos is very steep. On the other hand, the maximum energy depends more sensitive to the distance for the non-thermal neutrinos; indeed, we find that all detectors are capable of capturing neutrinos with $> 100$ MeV if the source is located at $\lesssim 4$ kpc. It should be stressed that HK will detect $> 100$ MeV neutrinos for CCSNe with $\lesssim 10$ kpc.

Towards real observations, we here suggest an appropriate strategy for the analysis of high energy neutrinos in the early post bounce phase. For a Galactic CCSN, we will catch a signal of neutronization burst, which will enable us to make a rough estimation of a time of core bounce. We then count up the neutrino detection and analyze them with the duration of $\Delta T \lesssim 50$ ms. It should be noted that the duration of the time window seems to be important to determine the physical parameters of non-thermal neutrinos, since the information on the non-thermal component may be smeared out by the thermal one if the duration is too long. This is due to the fact that the shock acceleration only occurs in the limited phase (see Sec.~\ref{sec:theoryshockac} for more details), meanwhile the thermal emission of neutrinos is persistent until the PNS completely cools off by neutrinos or it is engulfed by a BH.

\subsection{Detectability of high energy neutrinos from failed CCSN}\label{subsec:deteBHcase}
\begin{figure*}
\centering
\includegraphics[width=1.0\textwidth]{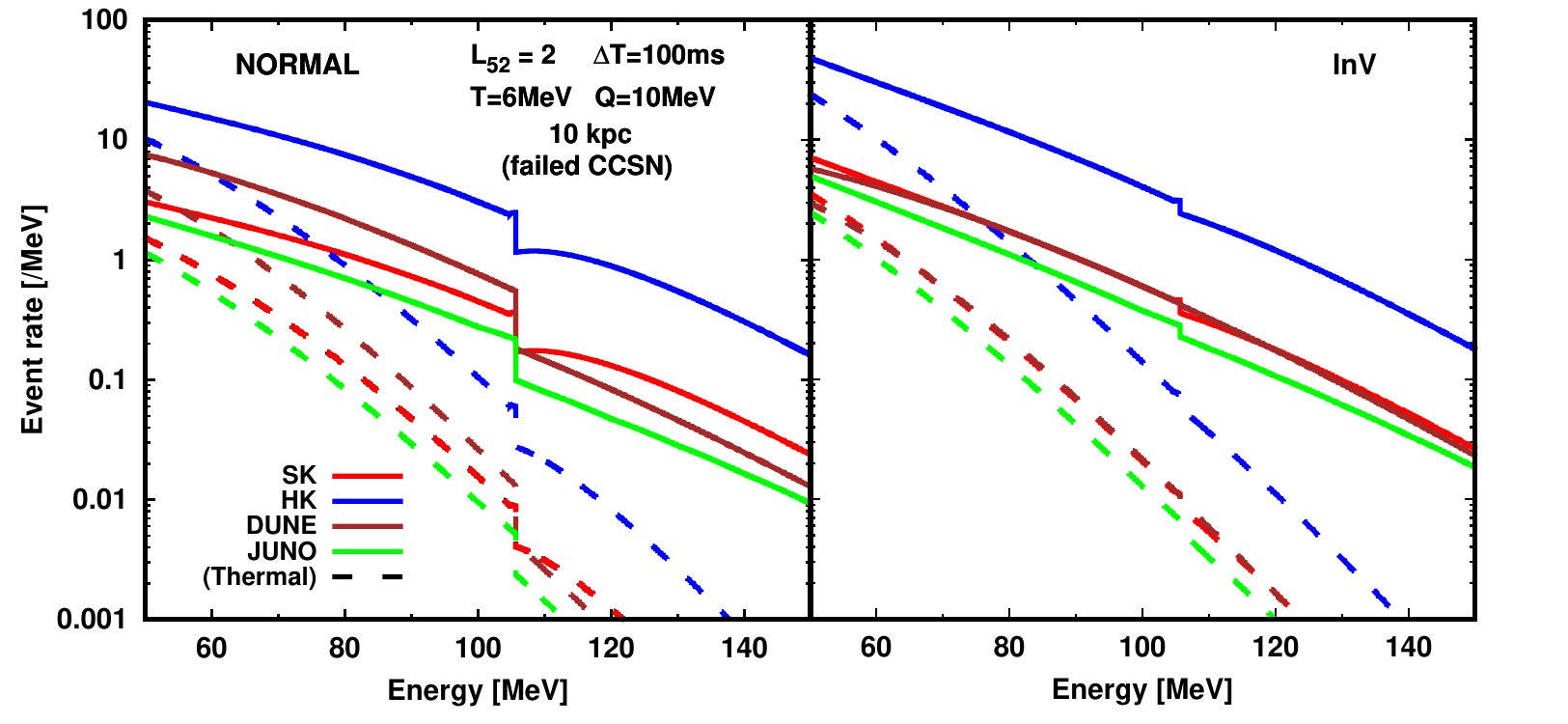}
\caption{Same as Fig.~\ref{graph_Detectionnumspectrum} but for the late post bounce phase in failed CCSN.}\label{graph_Detectionnumspectrum_BH}
\end{figure*}

\begin{figure*}
\centering
\includegraphics[width=1.0\textwidth]{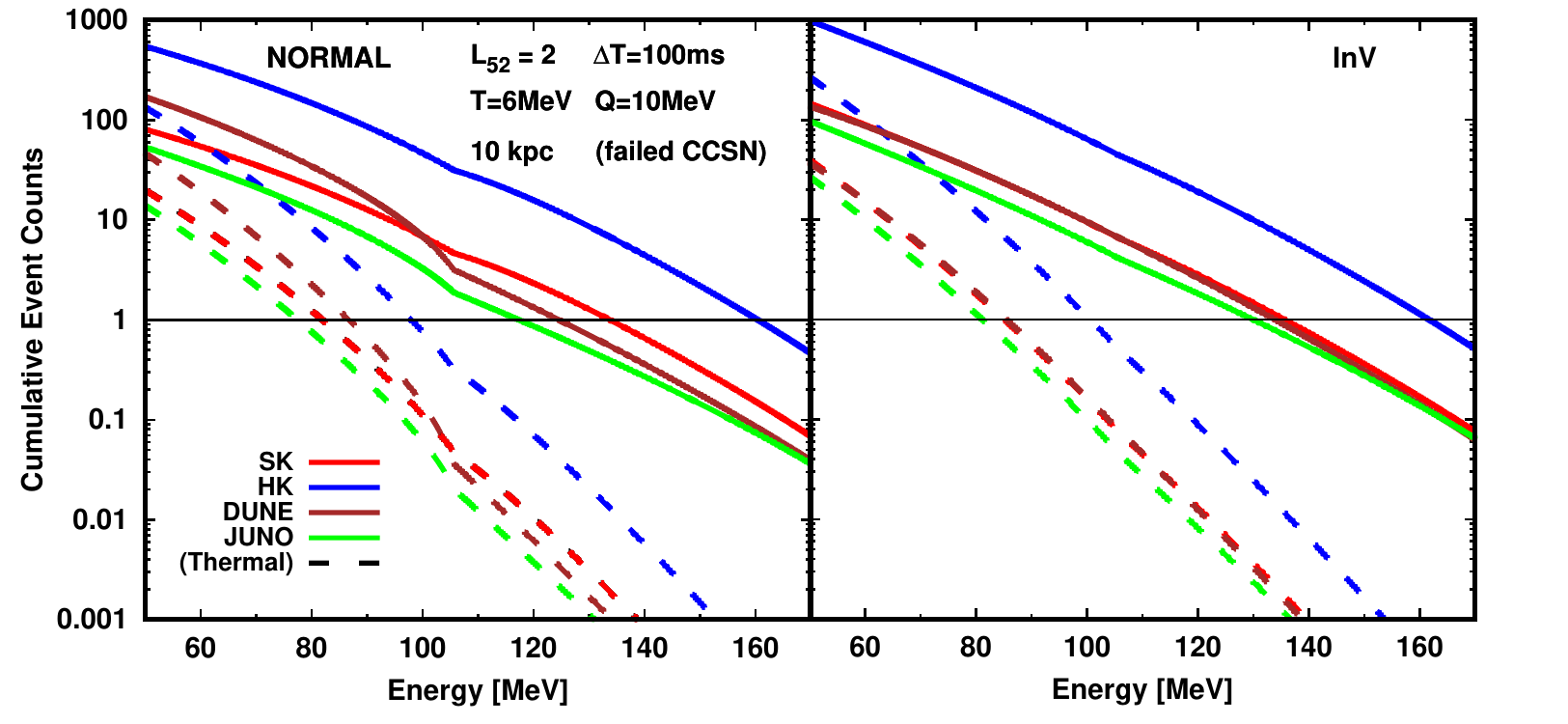}
\caption{Same as Fig.~\ref{graph_CumulativeFromHigh} but for the late post bounce phase in failed CCSN.}\label{graph_CumulativeFromHigh_BH}
\end{figure*}

We now turn our attention to the late post bounce phase in the case with failed CCSN. The energy spectra of event counts and those of the cumulative counts on each detector are shown in Fig.~\ref{graph_Detectionnumspectrum_BH} and \ref{graph_CumulativeFromHigh_BH}, respectively. In general, the event counts are expected to be higher than that in the early post bounce phase, which are mainly due to two reasons. One is that the temperature of thermal neutrino in the late post bounce phase is higher than that in the early one. We note that the thermal neutrinos play an important role to determine the overall scale of the non-thermal neutrinos. As described in Sec.~\ref{sec:theoryshockac}, the origin of the non-thermal neutrinos is the thermal component at $E \sim 50$ MeV where the spectrum is quasi-thermal (see also Eq.~\ref{eq:epsiBrela} in the relation between thermal- and non-thermal neutrinos in our model). The spectrum shape is an exponential decline around the energy, implying that the number of injected neutrinos is sensitive to the temperature. The other reason is that the duration time of neutrino shock acceleration is $10$ times longer than that in the early post-bounce phase.

 As shown in Fig.~\ref{graph_CumulativeFromHigh_BH}, all the detectors which we employ in this study will detect high energy neutrinos with $E > 100$ MeV if the source is located at $10$ kpc. It should be stressed that the event counts by thermal neutrinos are orders of magnitude smaller than the total for neutrinos, indicating that the detection of neutrinos with $E > 100$ MeV remains a smoking gun evidence that the neutrino shock acceleration occurs in CCSNe. Our results also suggest that the detection of high energy neutrinos from Galactic failed CCSN is feasible, and it would distinguish a BH formation by a failed CCSN from that by fall-back accretions in successful CCSN, since no high energy neutrinos in the late phase can be expected in the latter case. The parameters to characterize the non-thermal neutrinos will be constrained accurately in this case due to higher statistics than that in the early phase. It should be also mentioned that the expected maximum energy of detected neutrinos is above $150$ MeV for HK, indicating that there is a possibility to identify $\nu_{\mu}$ and $\bar{\nu}_{\mu}$ events in detectors by Cherenkov lights emitted by muons. Let us move on to the discussion regarding muon productions.

\subsection{Muon production in detectors}\label{subsec:muondetec}
The $\nu_{\mu}$ and $\bar{\nu}_{\mu}$ with $E > M_u$ in detectors potentially create muons through charged current reactions with detector materials. If it happens by CCSN neutrinos in reality, this will be a precious information to study the neutrino oscillation. In this section, we estimate the event count on each detector based on our models of the neutrino shock acceleration, and then discuss the necessary conditions to observe them.

\begin{figure*}
\centering
\includegraphics[width=1.0\textwidth]{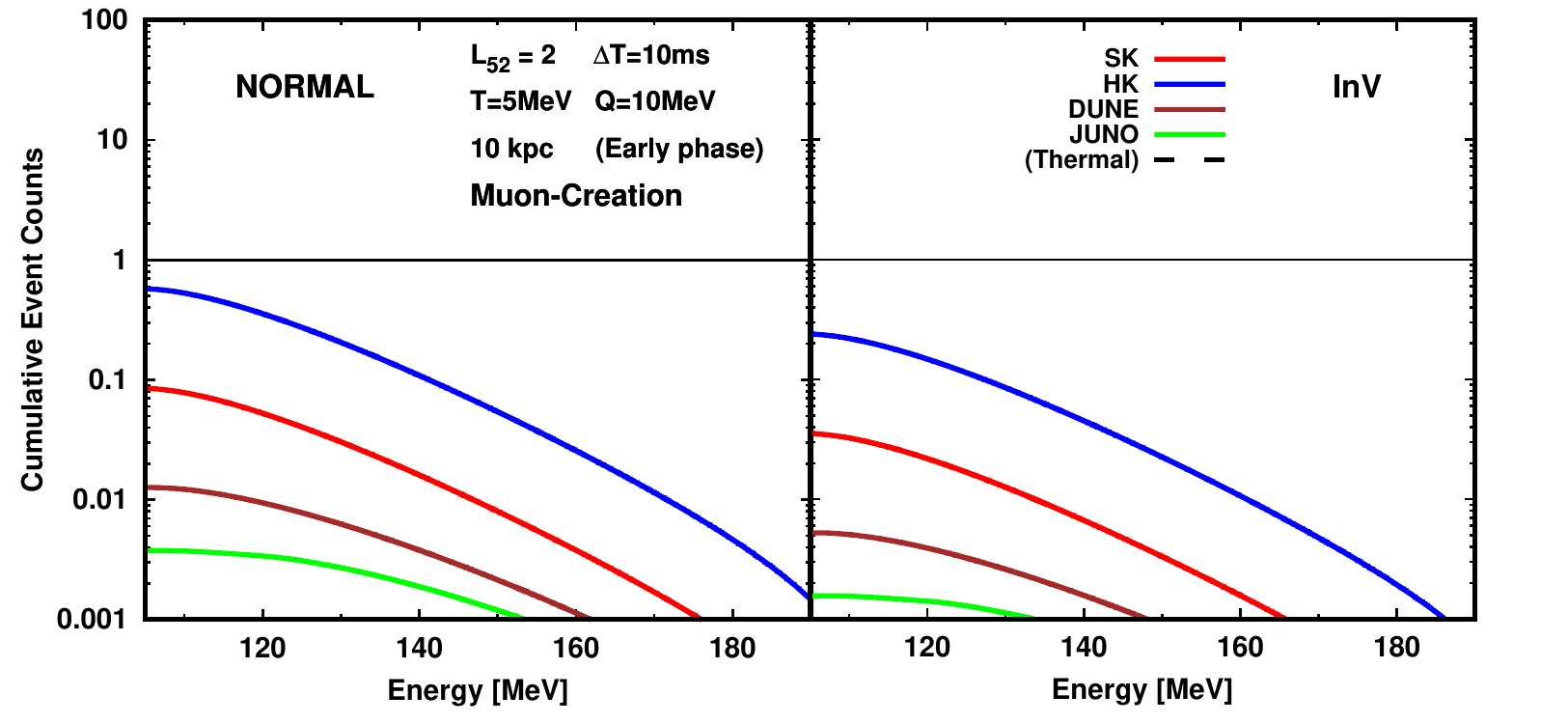}
\caption{Same as Fig.~\ref{graph_CumulativeFromHigh} but only for charged current reaction channels with muons.}\label{graph_CumulativeFromHigh_muon}
\end{figure*}

\begin{figure*}
\centering
\includegraphics[width=1.0\textwidth]{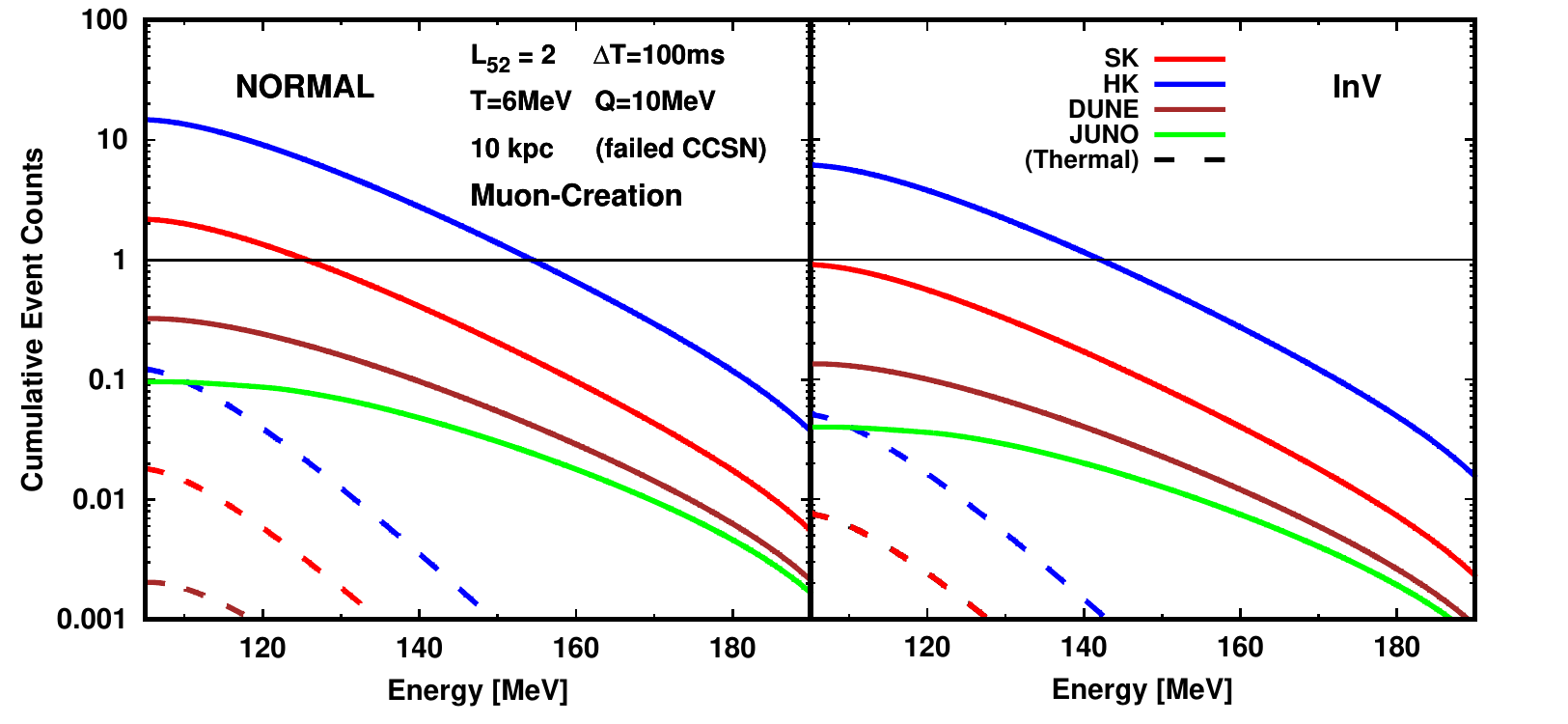}
\caption{Same as Fig.~\ref{graph_CumulativeFromHigh_BH} but only for charged current reaction channels with muons.}\label{graph_CumulativeFromHigh_BH_muon}
\end{figure*}

Fig.~\ref{graph_CumulativeFromHigh_muon} shows the cumulative number of events relevant to charged-current reactions with $\nu_{\mu}$ and $\bar{\nu}_{\mu}$ in the early post bounce phase. Unfortunately, the expected event counts are less than 1 for all detectors, indicating that the muon production may not happen in this case. It should be noted, however, there remains a possibility to detect them, in particular for HK, by taking into account uncertainties of the parameters and the neutrino cross sections. The distance to the CCSN source is also another important factor to discuss the detectability, which will be discussed below in more details. On the other hand, there are higher possibilities to produce muons in the late post-bounce phase for failed CCSN, which can be seen in Fig.~\ref{graph_CumulativeFromHigh_BH_muon}. As shown in the plot, $\sim 10$ muons may be produced in HK, and a few of them have an energy of $E \gtrsim 150$ MeV. Such high energy neutrinos will produce muons with large enough kinetic energy to emit Cherenkov lights in detectors. It should be also mentioned that muon production may occur in SK, although more quantitative study is necessary for the assessment of the detectability. We find that the charged current reactions with Oxygen play a dominant role for the muon productions for the Water Cherenkov detectors.

For our reference parameters, muon productions in DUNE and JUNO seem to be unlikely if the CCSN distance is $\geq 10$ kpc. As discussed in Sec.~\ref{sec:MonteCarlo}, however, the efficiency of neutrino shock acceleration depends on the mass accretion rate in the late post bounce. This indicates that the parameter $Q$, which characterizes the exponential decline of non-thermal spectrum, can be higher than our reference value ($Q=10$ MeV). There is the other positive effect that temperature of the thermal component would be higher in real than our reference value ($T = 6$ MeV); hence, it would be premature to conclude the detectability for muon productions. More detailed study should be made with more quantitative arguments, although it is dispensable to reduce the uncertainties of neutrinos cross sections with heavy nuclei, as repeatedly mentioned in this paper.

\begin{figure*}
\centering
\includegraphics[width=0.8\textwidth]{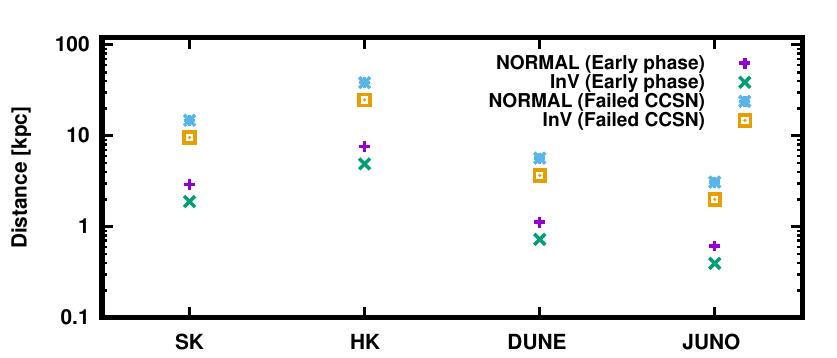}
\caption{The threshold distance to the source for which the high energy neutrinos create muons on each detector. Each symbol distinguishes the phase (early phase or late phase with BH formation) and the neutrino mass hierarchy.}\label{graph_MuonDetecDistance}
\end{figure*}

Fig.~\ref{graph_MuonDetecDistance} displays the threshold distance to CCSN where the emitted neutrinos can produce muons on each detector. In the early post bounce phase, the muon production requires that the CCSN is located nearby ($\lesssim 5$ kpc and $\lesssim 3$ kpc, respectively) for HK and SK or very nearby ($\lesssim 1$ kpc and $\lesssim 0.5$ kpc, respectively) for DUNE and JUNO. In the late phase for failed CCSN, on the other hand, the threshold distance is increased by a factor of $\sim 4$ than that in the early phase, indicating muon productions likely occur in HK for all Galactic failed CCSN. We also find an interesting trend that the threshold distance for the case with normal mass hierarchy is systematically larger than that for inverted mass hierarchy. This is attributed to the fact that most of $\nu_{\tau}$ at the CCSN source arrive at the Earth as $\nu_e$ in inverted mass hierarchy (see the spectrum at $E > M_u$ in Fig.~\ref{graph_neutrinospectrum_inv}), which results in the substantial reduction of $\nu_{\mu}$ events in terrestrial detectors. In normal mass hierarchy, on the other hand, roughly 3 flavors of neutrinos at the Earth share the original $\nu_{\tau}$ at the CCSN source; hence, the event rate of muon production becomes higher than that in inverted one.

\section{Summary and conclusion}\label{sec:sum}
The energy spectrum of neutrinos from CCSN have been regarded as a quasi-thermal shape, which seems to be real for most of the phases. As pointed out by the earlier studies in \citet{Kazanas1981ICRC,Giovanoni1989ApJ}, however, that the neutrino shock acceleration occurs in the early post bounce phase, which potentially creates the non-thermal shape in the emergent spectrum. This argument is also supported by recent CCSN simulations with full Boltzmann neutrino transport (see Fig.~\ref{graph_fout_nux_radialdistri_25ms_forpaper}). We also speculate that the neutrino shock acceleration occurs in the late post bounce phase for failed CCSN, in which the PNS is enveloped by a stalled shock wave located at $\sim 100$ km with high mass accretion rate. Motivated by these considerations, we perform a comprehensive study of the neutrino shock acceleration from the production mechanism to their detectability by representative terrestrial neutrino observatories. The main conclusions are summarized below.

\begin{enumerate}
\item The neutrino shock acceleration is strongly flavor dependent; $\nu_{\tau}$ (and $\bar{\nu}_{\tau}$) gains the energy up to $\sim 200$ MeV, meanwhile $\nu_{\mu}$ (and $\bar{\nu}_{\mu}$) has the similar spectrum as that in $\nu_{\tau}$ but sharp cut-off would appear at the energy of ($\sim 100$ MeV). The spectrum for both $\nu_{e}$ and $\bar{\nu}_e$ remains a quasi-thermal shape, since the shock acceleration is hampered by their charged current reactions with nucleons.
\item The observable non-thermal neutrinos need to satisfy the condition of Eq.~\ref{eq:tau}, otherwise neutrinos escape from the shock wave without interacting to the shock wave or advect with the accretion flows (see Sec.~\ref{sec:theoryshockac} for more details).
\item We demonstrate the neutrino shock acceleration by employing a Monte Carlo neutrino transport simulations and then obtain the emergent spectra. We also study how the acceleration efficiency changes with time in early post bounce phase and how it depends on the mass accretion rate in the late phase for failed CCSN. See Sec.~\ref{sec:MonteCarlo} for more details.
\item We assess the detectability of high energy neutrinos generated by the shock acceleration in four representative neutrino detectors for CCSN neutrinos. In this study, we adopt a simple but three flavor neutrino oscillation model (adiabatic MSW model). We find that the maximum energy of observable neutrinos is remarkably different from that of thermal ones (see Fig.~\ref{graph_Detectionthreshold} for the early post bounce phase). It should be also mentioned that the event counts above $\sim 80$ MeV becomes a few orders of magnitude higher than that of the thermal neutrinos regardless of any detectors, indicating that the detection of such a high energy neutrinos will provide an evidence of neutrino shock acceleration (see Secs.~\ref{subsec:deteearlyresults}~and~\ref{subsec:deteBHcase} for more details).
\item Our result suggests that the high energy neutrinos from failed CCSN are abundantly produced in the late post bounce phase ($\gtrsim 100$ ms). If the detectors catch the signal, it will distinguish a BH formation by a failed CCSN from that by fall-back accretions in successful CCSN (see Sec.~\ref{subsec:deteBHcase} for more details).
\item The detection of the high energy neutrinos through charged current reactions will be a smoking gun evidence that neutrinos experience flavor conversions. In connection with this argument, the neutrino shock acceleration opens up a new possibility of muon productions in detectors. If muons are observed in the detectors, this will be a precious information to place a constraint of the physics of neutrino oscillation (see Sec.~\ref{subsec:muondetec} for more details).
\end{enumerate}

We must mention some caveats that need to be addressed in future. First, although we do not take into account reaction channels with neutral current in detectors, they would play important roles for the data analysis in reality. Since the reactions are sensitive to all flavor of neutrinos, they provide a number spectrum of flavor-integrated neutrinos. By combining the data with charged current reactions, it will place a constraint on transition probabilities from each heavy leptonic neutrino to other species. We also note that the rapid decline of neutrino number flux at $E \sim 100$ MeV would appear more clearly in neutral current reactions than that of the charged current, indicating that the disappearance of $\nu_{\mu}$ and $\bar{\nu}_{\mu}$ around the energy at the CCSN source will be directly proven. Those demonstrations are currently undertaken and will be published with more detailed arguments in forthcoming papers.

Second, although our Monte Carlo simulations provide representative examples of emergent neutrino spectra generated by the shock acceleration, they depend on the input physics (in particular neutrino-matter interactions) and also matter dynamics. This indicates that the systematic study of the parameter dependence in our analytic formula is necessary for more quantitative discussions, although the reduction of uncertainties in neutrino cross sections with detector materials is indispensable. The survival probabilities of each flavor of neutrinos is also another important parameter for the estimation, which should be improved with more realistic neutrino oscillation model (for instance, including the Earth matter effect and collective neutrino oscillations).

Third, we point out another possible case that occurs the neutrino shock acceleration; if the QCD phase transition happens inside PNS, it would generate the secondary shock wave \citep[see, e.g.,][]{2010PhRvD..81j3005D,2020ApJ...894....9F}. Since the shock wave transits from the optically thick to thin regimes for neutrinos, the neutrino shock acceleration would occur similar as that in the early post bounce phase. This is also interesting and deserves to be considered in more details.

As a concluding but an important remark, we find that the event counts of the high energy neutrinos would not be large, despite of the fact that the neutrino shock acceleration increases them by a few orders of magnitude. This indicates that the detection statistics on each detector will be low. Hence, it will be a crucial to share the observed data of each neutrino detector to develop reliable analyses of high energy neutrinos. The joint analysis will enable us to see a flavor-dependent feature in CCSN neutrinos and place a constraint on the physics of neutrino oscillation. We hope that our study provides a springboard for the new collaboration in the CCSN and experimental communities.

\section*{Acknowledgements}
We are grateful to Kate Scholberg, Joel Dai, and Yudai Suwa for useful comments and discussions. We acknowledge support from the U.S. Department of Energy Office of Science and the Office of Advanced Scientific Computing Research via the Scientific Discovery through Advanced Computing (SciDAC4) program and Grant DE-SC0018297 (subaward 00009650).

\section*{DATA AVAILABILITY}
The data underlying this article will be shared on reasonable request to the corresponding author.

\section*{Appendix}
The hydrodynamic quantities are described by the following analytic functions, which approximate the results in the early post-bounce phase ($\sim 30\,{\rm ms}$) in a CCSN simulation of \citet{2019ApJS..240...38N}. We note that the analytic treatment is more useful than the numerical data for the post-processing simulations, since the results may be affected by the time-interval of data-output of CCSN simulations. In our analytic prescription, the time evolution of the system can be simply characterized by the shock radius. The shock velocity is estimated from the result of a CCSN simulation.

The number densities of nucleons in post shock region, $\mbox{$r\leq r_{\rm sh}$}$, and nuclei in pre-shock region, $\mbox{$r> r_{\rm sh}$}$, are given by
\begin{eqnarray}
n(r,t) = 
\left\{ \begin{array}{ll}
 n_0\left(\frac{r}{r_0}\right)^{-4.4}\,{\rm cm^{-3}} 
 &\mbox{$r\leq r_{\rm sh}$}, \\
n_0^{\prime}\left(\frac{r_{\rm sh}}{r_0}\right)^{-4.4}\left(\frac{r}{r_{\rm sh}}\right)^{-2}\,{\rm cm^{-3}}
 &\mbox{ $r>r_{\rm sh}$},
       \end{array} \right. \label{eq:n}
\end{eqnarray}
where we set $n_0=1.4\cdot 10^{12}\,{\rm cm^{-3}}$, $n_0^{\prime}=n_0/\langle A_{\rm nuc}\rangle/10$, $r_0=4\cdot 10^6\,{\rm cm}$, and $\langle A_{\rm nuc}\rangle$ is the mean nuclear mass number in the upstream. {As shown in \citet{2019ApJS..240...38N}, the photodissociation of heavy nuclei may be incomplete at the shock wave, implying that a substantial light nuclei would appear in right behind the shock wave. In this case, the dominant energy loss of neutrinos in post shock region may not be the recoil of nucleons but that of light nuclei, implying that the maximum energy of accelerated $\nu_{\tau}$ may exceed $200$ MeV. Although this is an interesting possibility, we postpone the actual impact of light nuclei in neutrino shock acceleration in future work.}

The velocity field is given by
\begin{eqnarray}
v(r,t) \approx 
\left\{ \begin{array}{ll}
 -v_{\rm max}/10 &\mbox{$r\leq r_{\rm sh}$}, \\
- v_{\rm max}\left(\frac{r}{r_{\rm sh}}\right)^{-1/2}
 &\mbox{ $r>r_{\rm sh}$},
       \end{array} \right. \label{eq:v}
\end{eqnarray}
where $v_{\rm max}=0.2c$.
The temperature is 
\begin{eqnarray}
T(r,t) \approx 
\left\{ \begin{array}{ll}
 12.8\left(\frac{r}{23\,{\rm km}}\right)^{-1.2}\,{\rm MeV} &\mbox{$r\leq r_{\rm sh}$}, \\
T_0\left(\frac{r}{r_{\rm sh}}\right)^{-0.3}\,{\rm MeV}
 &\mbox{ $r>r_{\rm sh}$},
       \end{array} \right. \label{eq:T}
\end{eqnarray}
where $T_0=1.2(r_{\rm sh}/75\,{\rm km})^{-0.7}\,{\rm MeV}$.
The electron fraction is described by
\begin{eqnarray}
Y_e(r,t) \approx 
\left\{ \begin{array}{ll}
{\rm min}(y,0.5) &\mbox{$r\leq r_{\rm sh}$}, \\
0.5
 &\mbox{ $r>r_{\rm sh}$},
       \end{array} \right. \label{eq:v}
\end{eqnarray}
where 
\begin{eqnarray}
y & = & (y_1^s+y_2^s+y_3^s)^{1/s},\\
y_1 & = & 0.15\left(\frac{r}{r_{\rm sh}}\right)^{0.3},\\
y_2 & = & 0.49\left(\frac{r}{r_{\rm sh}}\right)^{7},\\
y_3 & = & 0.28\left(\frac{r}{22\,{\rm km}}\right)^{-1.5}.\\
\end{eqnarray}
Here we choose $s=3$.
We describe the electron chemical potential as
\begin{eqnarray}
\mu_e(r,t) \approx 
\left\{ \begin{array}{ll}
(\mu_1^s+\mu_2^s)^{1/s} &\mbox{$r\leq r_{\rm sh}$}, \\
6\,{\rm MeV}
 &\mbox{ $r>r_{\rm sh}$},
       \end{array} \right. \label{eq:v}
\end{eqnarray}
where 
\begin{eqnarray}
\mu_1 & = & 55\left(\frac{r}{20\,{\rm km}}\right)^{-1.3}\,{\rm MeV},\\
\mu_2 & = & 55\left(\frac{r_{\rm sh}}{29\,{\rm km}}\right)^{-1.3}\,{\rm MeV}.
\end{eqnarray}
Although we set the temperature, electron fraction, and electron chemical potential in the upstream, we can neglect electron scattering in this region. Thus, the choices of the parameters for the upstream does not affect our result.





\bibliographystyle{mnras}
\bibliography{bibfileKH,bibfileHN}







\bsp	
\label{lastpage}
\end{document}